\documentclass[12pt,prd,onecolumn,showpacs,amsmath,amssymb,aps,floats,floatfix,nofootinbib]{revtex4-1}

\usepackage[colorlinks=true,urlcolor=blue,anchorcolor=blue,citecolor=blue,filecolor=blue,linkcolor=blue,menucolor=blue,linktocpage=true]{hyperref} % should be commented out if the tex file will be compiled with latex in arXiv!!! (pdflatex is fine)

\usepackage[inline]{enumitem}
\usepackage{dcolumn}
\usepackage{bm}
\usepackage{amsmath}
\usepackage{amsfonts}
\usepackage{amssymb}
\usepackage{color}
\usepackage{float}
\usepackage{latexsym}
\usepackage{slashed} % slash mark
\usepackage{pstricks}
\usepackage{indentfirst}
\usepackage{mathrsfs}
\usepackage{multirow}
\usepackage{epsfig,psfrag}
\usepackage{subfigure}
\usepackage{mathtools}
\usepackage{setspace} % spacing
\usepackage[utf8]{inputenc} % accept utf-8 input encoding
\usepackage[scientific-notation=true]{siunitx} % comprehensive units

\graphicspath{{fig/}}

\allowdisplaybreaks % allow eqnarray breaks
\begin{document}

\title{Fermionic and scalar dark matter with hidden $\mathrm{U}(1)$ gauge interaction and kinetic mixing}
\author{Juebin Lao}
\author{Chengfeng Cai}
\author{Zhao-Huan Yu}\email{yuzhaoh5@mail.sysu.edu.cn}
\author{Yu-Pan Zeng}
\author{Hong-Hao Zhang}\email{zhh98@mail.sysu.edu.cn}
\affiliation{School of Physics, Sun Yat-Sen University, Guangzhou 510275, China}

\begin{abstract}

We explore the Dirac fermionic and complex scalar dark matter in the framework of a hidden $\mathrm{U}(1)_\mathrm{X}$ gauge theory with kinetic mixing between the $\mathrm{U}(1)_\mathrm{X}$ and $\mathrm{U}(1)_\mathrm{Y}$ gauge fields. The $\mathrm{U}(1)_\mathrm{X}$ gauge symmetry is spontaneously broken due to a hidden Higgs field. The kinetic mixing provides a portal between dark matter and standard model particles. Besides, an additional Higgs portal can be realized in the complex scalar case. Dark matter interactions with nucleons are typically isospin violating, and direct detection constraints can be relieved. Although the kinetic mixing has been stringently constrained by electroweak oblique parameters, we find that there are several available parameter regions predicting an observed relic abundance through the thermal production mechanism. Moreover, these regions have not been totally explored in current direct and indirect detection experiments.
Future direct detection experiments and searches for invisible Higgs decays at a Higgs factory could further investigate these regions.

\end{abstract}

\maketitle
\tableofcontents
\clearpage

\section{Introduction}

The standard model (SM) with $\mathrm{SU}(3)_\mathrm{C} \times \mathrm{SU}(2)_\mathrm{L} \times \mathrm{U}(1)_\mathrm{Y}$ gauge interactions has achieved a dramatic success in explaining experimental data in particle physics.
Nonetheless, the SM must be extended for taking into account dark matter (DM) in the Universe, whose existence is established by astrophysical and cosmological experiments~\cite{Jungman:1995df,Bertone:2004pz,Feng:2010gw,Young:2016ala}.
The standard paradigm assumes dark matter is thermally produced in the early Universe, typically requiring some mediators to induce adequate DM interactions with SM particles.

Inspired by the gauge interactions in the SM, it is natural to imagine dark matter participating a new kind of gauge interaction.
The simplest attempt is to introduce an additional $\mathrm{U}(1)_\mathrm{X}$ gauge symmetry with a corresponding gauge boson acting as a mediator~\cite{Langacker:2008yv}.
In order to minimize the impact on the interactions of SM particles, one can assume that all SM fields do not carry $\mathrm{U}(1)_\mathrm{X}$ charges~\cite{Feldman:2007wj,Pospelov:2007mp,Mambrini:2010dq,Kang:2010mh,Chun:2010ve,Mambrini:2011dw,Frandsen:2011cg,Gao:2011ka,Chu:2011be,Frandsen:2012rk,Jia:2013lza,Belanger:2013tla,Chen:2014tka,Arcadi:2017kky,Liu:2017lpo,Dutra:2018gmv,Bauer:2018egk,Koren:2019iuv,Jung:2020ukk}.
Thus, such a $\mathrm{U}(1)_\mathrm{X}$ gauge interaction belongs to a hidden sector, which also involves dark matter and probably an extra Higgs field generating mass to the $\mathrm{U}(1)_\mathrm{X}$ gauge boson via the Brout-Englert-Higgs mechanism~\cite{Higgs:1964ia,Higgs:1964pj,Englert:1964et}\footnote{The Stueckelberg mechanism~\cite{Stueckelberg:1900zz,Chodos:1971yj} is another way to generate the gauge boson mass.}.
It is easy to make the theory free from gauge anomalies by assuming the DM particle is a Dirac fermion or a complex scalar boson.
Gauge symmetries allow a renormalizable kinetic mixing term between the $\mathrm{U}(1)_\mathrm{X}$ and $\mathrm{U}(1)_\mathrm{Y}$ field strengths~\cite{Holdom:1985ag}, which provides a portal connecting DM and SM particles.

In this paper, we focus on DM models with a hidden $\mathrm{U}(1)_\mathrm{X}$ gauge symmetry, which is spontaneously broken due to a hidden Higgs field.
We assume that the DM particle is a $\mathrm{SU}(3)_\mathrm{C} \times \mathrm{SU}(2)_\mathrm{L} \times \mathrm{U}(1)_\mathrm{Y}$ gauge singlet but carries a $\mathrm{U}(1)_\mathrm{X}$ charge.
Because of the kinetic mixing term, the $\mathrm{U}(1)_\mathrm{X}$ and $\mathrm{U}(1)_\mathrm{Y}$ gauge fields mix with each other, modifying the electroweak oblique parameters $S$ and $T$ at tree level~\cite{Holdom:1990xp,Babu:1997st}.
In the mass basis, electrically neutral gauge bosons include the photon, the $Z$ boson, and a new $Z'$ boson.
The $Z$ and $Z'$ bosons couple to both the DM particle and SM fermions, based on the kinetic mixing portal.
As a result, DM couplings to protons and neutrons are typically different~\cite{Kang:2010mh,Frandsen:2011cg,Gao:2011ka,Chun:2010ve,Belanger:2013tla,Chen:2014tka}, leading to isospin-violating DM-nucleon scattering~\cite{Feng:2011vu} in direct detection experiments.

In this framework, specifying different spins of the DM particle and various $\mathrm{U}(1)_\mathrm{X}$ charges in the hidden sector would lead to different DM models.
The simplest case is to consider Dirac fermionic DM, whose phenomenology has been studied in Refs.~\cite{Mambrini:2010dq,Chun:2010ve,Liu:2017lpo,Bauer:2018egk}.
Firstly, we revisit this case, investigating current constraints from electroweak oblique parameters, DM relic abundance, and direct and indirect detection experiments.
Nonetheless, it is not easy to accommodate the constraints from relic abundance and direct detection, except for some specific parameter regions.
The main reason is that DM annihilation in the early Universe due to the kinetic mixing portal alone is generally too weak, tending to overproduce dark matter.

Therefore, we go further to consider the case of complex scalar DM, which could have quartic couplings to both the SM and hidden Higgs fields.
Consequently, the DM particle can also communicate with the SM fermions mediated by two Higgs bosons, which are mass eigenstates mixed with the SM and hidden Higgs bosons.
Such an additional Higgs portal can help enhance DM annihilation.
Moreover, it can also adjust the DM-nucleon couplings and weaken the direct detection constraint.
Thus, it should be easier to find viable parameter regions in the complex scalar DM case.

This paper is organized as follows.
In Sec.~\ref{sec:U1X_gauge}, we review the hidden $\mathrm{U}(1)_\mathrm{X}$ gauge theory with kinetic mixing and study the constraint from electroweak oblique parameters. 
In Secs.~\ref{sec:fermionic_DM} and \ref{sec:scalar_DM}, we discuss a Dirac fermionic DM model and a complex scalar DM model, respectively,
and investigate the constraints from the relic abundance observation, and direct and indirect detection experiments.
Finally, we give the conclusions and discussions in Sec.~\ref{sec:concl}.

\section{Hidden $\mathrm{U}(1)_\mathrm{X}$ gauge theory}
\label{sec:U1X_gauge}

In this section, we briefly review the hidden $\mathrm{U}(1)_\mathrm{X}$ gauge theory with the kinetic mixing between the $\mathrm{U}(1)_\mathrm{X}$ and $\mathrm{U}(1)_\mathrm{Y}$ gauge fields.
Furthermore, we investigate the constraints from electroweak oblique parameters.

\subsection{Hidden U(1)$_\mathrm{X}$ gauge theory with kinetic mixing}

We denote the $\mathrm{U}(1)_\mathrm{Y}$ and $\mathrm{U}(1)_\mathrm{X}$ gauge fields as $\hat{B}_\mu$ and $\hat{Z}'_\mu$, respectively.
Their gauge invariant kinetic terms in the Lagrangian reads
\begin{eqnarray}\label{kin_mix}
\mathcal{L}_{\mathrm{K}} &=&
-\frac{1}{4} \hat{B}^{\mu\nu}\hat{B}_{\mu\nu}
-\frac{1}{4} \hat{Z}'^{\mu\nu}\hat{Z}'_{\mu\nu}
-\frac{s_\varepsilon}{2} \hat{B}^{\mu\nu}\hat{Z}'_{\mu\nu}
\nonumber\\
&=& -\frac{1}{4}
\begin{pmatrix}
\hat{B}^{\mu\nu}, & \hat{Z}'^{\mu\nu}
\end{pmatrix}
\begin{pmatrix}
1 & s_\varepsilon\\
s_\varepsilon & 1
\end{pmatrix}
\begin{pmatrix}
\hat{B}_{\mu\nu}\\
\hat{Z}'_{\mu\nu}
\end{pmatrix},
\end{eqnarray}
where the field strengths are $\hat{B}_{\mu\nu} \equiv \partial_\mu \hat{B}_\nu - \partial_\nu \hat{B}_\mu$ and $\hat{Z}'_{\mu\nu} \equiv \partial_\mu \hat{Z}'_\nu - \partial_\nu \hat{Z}'_\mu$.
The $s_\varepsilon$ term is a kinetic mixing term, which makes the kinetic Lagrangian \eqref{kin_mix} in a noncanonical form.
Achieving correct signs for the diagonalized kinetic terms requires $s_\varepsilon \in (-1,1)$.
Thus, we can define an angle $\varepsilon \in (-\pi/2,\pi/2)$ satisfying $s_\varepsilon \equiv \sin\varepsilon$.
The kinetic Lagrangian \eqref{kin_mix} can be made canonical via a $\mathrm{GL}(2,\mathbb{R})$ transformation~\cite{Babu:1997st},
\begin{equation}
V_\mathrm{K} = \begin{pmatrix}
1 & -t_\varepsilon \\
0 & 1/c_\varepsilon
\end{pmatrix},
\end{equation}
which satisfies
\begin{equation}
V_\mathrm{K}^\mathrm{T}
\begin{pmatrix}
1 & s_\varepsilon\\
s_\varepsilon & 1
\end{pmatrix}
V_\mathrm{K}
= \setlength{\arraycolsep}{.4em}
\begin{pmatrix}
1 & \\
 & 1
\end{pmatrix}.
\end{equation}
Here we have adopted the shorthand notations $c_{\varepsilon}\equiv \cos\varepsilon $ and $t_{\varepsilon}\equiv \tan\varepsilon$.

We assume that the $\mathrm{U}(1)_\mathrm{X}$ gauge symmetry is spontaneously broken by a hidden Higgs field $\hat{S}$ with $\mathrm{U}(1)_\mathrm{X}$ charge $q_S = 1$.
Now the Higgs sector involves $\hat{S}$ and the SM Higgs doublet $\hat{H}$. The corresponding Lagrangian respecting the $\mathrm{SU}(2)_\mathrm{L} \times \mathrm{U}(1)_\mathrm{Y} \times \mathrm{U}(1)_\mathrm{X}$ gauge symmetry reads~\cite{Liu:2017lpo}
\begin{eqnarray}\label{modelLagarange}
\mathcal{L}_{\mathrm{H}} &=& (D^\mu \hat{H})^\dag(D_\mu \hat{H})+(D^\mu \hat{S})^\dag(D_\mu \hat{S}) +\mu^2|\hat{H}|^2 +\mu_S^2
|\hat{S}|^2
\nonumber\\
&& -\frac{1}{2}\lambda_H|\hat{H}|^4 -\frac{1}{2}\lambda_S|\hat{S}|^4 -\lambda_{HS}|\hat{H}|^2|\hat{S}|^2.
\end{eqnarray}
The covariant derivatives are given by $D_\mu\hat{H}=(\partial_\mu-i\hat{g}'\hat{B}_\mu/2 -i\hat{g} W^a_\mu T^a)\hat{H}$ and $D_\mu\hat{S}=(\partial_\mu-ig_{X}\hat{Z}'_\mu)\hat{S}$,
where $W^a_\mu$ ($a=1,2,3$) denote the $\mathrm{SU}(2)_\mathrm{L}$ gauge fields and $T^a = \sigma^a/2$ are the $\mathrm{SU}(2)_\mathrm{L}$ generators.
$\hat{g}$, $\hat{g}'$, and $g_X$ are the corresponding gauge couplings.

Both $\hat{H}$ and $\hat{S}$ acquire nonzero vacuum expectation values (VEVs), $v$ and $v_S$, driving spontaneously symmetry breaking.
The Higgs fields in the unitary gauge can be expressed as
\begin{eqnarray}
\hat{H}&=&
\frac{1}{\sqrt{2}}
\begin{pmatrix}
0\\
v+H
\end{pmatrix},
\\
\hat{S}&=&
\frac{1}{\sqrt{2}} (v_S+S).
\end{eqnarray}
Vacuum stability requires the following conditions:
\begin{eqnarray}
\lambda _H > 0,\quad
\lambda_S > 0,\quad
\lambda_{HS} > -\sqrt{\lambda_H\lambda_S}\,.
\end{eqnarray}
The mass-squared matrix for $(H,S)$, 
\begin{equation}
\mathcal{M}_0^2 =
\begin{pmatrix}
\lambda_H v^2 & \lambda_{HS} v v_S\\
\lambda_{HS} v v_S & \lambda_S v_S^2
\end{pmatrix},
\end{equation}
can be diagonalized by a rotation with an angle $\eta$. The transformation between the mass basis $(h,s)$ and the gauge basis $(H,S)$ is given by
\begin{eqnarray}\label{eq:HS_to_hs}
\begin{pmatrix}
H\\
S
\end{pmatrix} &=&
\begin{pmatrix}c_\eta&-s_\eta\\
s_\eta&c_\eta
\end{pmatrix}
\begin{pmatrix}
h\\
s
\end{pmatrix},
\\
t_{2\eta}&=&\frac{2\lambda_{HS}vv_S}{\lambda_Hv^2-\lambda_Sv_S^2},
\label{t_2eta}
\end{eqnarray}
with the mixing angle $\eta \in [-\pi/4,\pi/4]$.
The physical masses of scalar bosons $h$ and $s$ satisfy
\begin{eqnarray}
m_h^2 &=& \frac{1}{2}\left[\lambda_H v^2 + \lambda_S v_S^2 + (\lambda _H v^2 - \lambda_S v_S^2)/c_{2\eta }\right],
\label{m_h_2}\\
m_s^2 &=& \frac{1}{2}\left[\lambda_H v^2 + \lambda_S v_S^2 + (\lambda _S v_S^2 - \lambda _H v^2)/c_{2\eta }\right].
\label{m_s_2}
\end{eqnarray}
Note that $h$ is the $125~\si{GeV}$ SM-like Higgs boson.
If $\lambda_{HS}$ vanishes, $h$ is identical to the SM Higgs boson.

The mass-squared matrix for the gauge fields $(\hat{B}_\mu, W^3_\mu, \hat{Z}'_\mu)$ generated by the Higgs VEVs reads
\begin{equation}
\mathcal{M}_1^2 =
\begin{pmatrix}
\hat{g}'^2 v^2/4 & -\hat{g}\hat{g}' v^2/4 & \\
-\hat{g}\hat{g}' v^2/4 & \hat{g}^2 v^2/4  & \\
 & & g_X^2 v_S^2
\end{pmatrix}.
\end{equation} 
Taking into account the kinetic mixing and the mass matrix diagonalization, the transformation between the mass basis $(A_\mu, Z_\mu, Z'_\mu)$ and the gauge basis $(\hat{B}_\mu, W^3_\mu, \hat{Z}'_\mu)$ is given by~\cite{Babu:1997st,Frandsen:2011cg}
\begin{equation}\label{rotation}
\begin{pmatrix}
\hat{B}_\mu\\
W^3_\mu\\ \hat{Z}'_\mu
\end{pmatrix}
=
V(\varepsilon) R_3(\hat{\theta}_W) R_1(\xi)
\begin{pmatrix}
A_\mu\\
Z_\mu\\
Z'_\mu
\end{pmatrix},
\end{equation}
with
\begin{eqnarray}
V(\varepsilon)&=&
\begin{pmatrix}
1& &-t_\varepsilon\\
 &1& \\
0& &1/c_\varepsilon
\end{pmatrix},
\\
R_3(\hat{\theta}_W)&=&
\begin{pmatrix}
\hat{c}_\mathrm{W}&-\hat{s}_\mathrm{W}& \\
\hat{s}_\mathrm{W}&\hat{c}_\mathrm{W} \\
 & &1\end{pmatrix},
\\
R_1(\xi)&=&
\begin{pmatrix}
1& & \\
 &c_\xi&-s_\xi\\
 &s_\xi&c_\xi
\end{pmatrix}.
\end{eqnarray}
Here, the weak mixing angle $\hat\theta_W$ satisfies
\begin{equation}
\hat{s}_\mathrm{W }\equiv \sin\hat\theta_W = \frac{\hat{g}'}{\sqrt{\hat{g}^2 + \hat{g}'^2}},\quad
\hat{c}_\mathrm{W }\equiv \cos\hat\theta_W = \frac{\hat{g}}{\sqrt{\hat{g}^2 + \hat{g}'^2}}.
\end{equation}
The rotation angle $\xi$ is determined by
\begin{eqnarray}\label{t2xi}
t_{2 \xi}=
\frac{s_{2\varepsilon}\hat{s}_\mathrm{W} v^{2}(\hat{g}^{2}+\hat{g}^{\prime 2})}
{c_{\varepsilon}^{2} v^{2}(\hat{g}^{2}+\hat{g}^{\prime 2})(1-\hat{s}_\mathrm{W}^{2} t_{\varepsilon}^{2})-4 g_{X}^{2} v_{S}^{2}}.
\end{eqnarray}
Note that $A_\mu$ and $Z_\mu$ correspond to the photon and $Z$ boson, and $Z'_\mu$ leads to a new massive vector boson $Z'$.
The photon remains massless, while the masses for the $Z$ and $Z'$ bosons are given by~\cite{Chun:2010ve}
\begin{eqnarray}
m_{Z}^{2}&=&\hat{m}_{Z}^{2}(1+\hat{s}_\mathrm{W} t_{\varepsilon} t_{\xi}),
\label{massofgauge:Z}\\
m_{Z^{\prime}}^{2}&=&\frac{\hat{m}_{Z^{\prime}}^{2}}{c_{\varepsilon}^{2}(1+\hat{s}_\mathrm{W} t_{\varepsilon} t_{\xi})},
\label{massofgauge:Zprime}
\end{eqnarray}
with $\hat{m}_{Z}^{2} \equiv (\hat{g}^{2}+\hat{g}^{\prime 2})v^{2}/4$ and $\hat{m}_{Z^{\prime}}^{2} \equiv g_{X}^{2} v_{S}^{2}$.
We define a ratio,
\begin{equation}\label{r}
r \equiv \frac{m_{Z'}^2}{m_{Z}^2},
\end{equation}
which will be useful in the following discussions.

The $W$ mass is $m_W = \hat{g}v/2$, only contributed by the VEV of $\hat{H}$, as in the SM.
Moreover, the charge current interactions of SM fermions at tree level are not affected by the kinetic mixing, remaining a form of
\begin{equation}
{\mathcal{L}_{{\mathrm{CC}}}} = \frac{1}{{\sqrt 2 }}(W_\mu ^ + J_W^{ + ,\mu } + \text{H.c.}),
\end{equation}
where the charge current is $J_W^{ + ,\mu } = {\hat g} ( {{\bar u}_{i{\mathrm{L}}}}{\gamma ^\mu }{V_{ij}}{d_{j{\mathrm{L}}}} + {{\bar \nu }_{i{\mathrm{L}}}}{\gamma ^\mu }{\ell _{i{\mathrm{L}}}})$ with $V_{ij}$ denoting the Cabibbo-Kobayashi-Maskawa matrix.
Consequently, the Higgs doublet VEV $v$ is still directly related to the Fermi constant $G_\mathrm{F} = \hat{g}^2/(4\sqrt{2}m_W^2) =(\sqrt{2}v^2)^{-1}$.

On the other hand, the neutral current interactions become
\begin{equation}
{\mathcal{L}_{{\mathrm{NC}}}} = j_{{\mathrm{EM}}}^\mu {A_\mu } + j_Z^\mu {Z_\mu } + j_{Z'}^\mu {Z'_\mu },
\end{equation}
where the electromagnetic current is $j_\mathrm{EM}^\mu  = \sum_f {{Q_f}e\bar f{\gamma ^\mu }f}$, with $e = \hat{g} \hat{g}' /\sqrt{\hat{g}^2 + \hat{g}^{'2}}$ and $Q_f$ denoting the electric charge of a SM fermion $f$.
The neutral current coupled to $Z$ is given by
\begin{equation}\label{j_Z_mu}
j_Z^\mu  = \frac{e{c_\xi }(1 + {{\hat s}_{\mathrm{W}}}{t_\varepsilon }{t_\xi })}{{2{{\hat s}_{\mathrm{W}}}{{\hat c}_{\mathrm{W}}}}}\sum\limits_f {\bar f{\gamma ^\mu }(T_f^3 - 2{Q_f}s_*^2 - T_f^3{\gamma _5})f}  + \frac{{{s_\xi }}}{{{c_\varepsilon }}} j^\mu_\mathrm{DM},
\end{equation}
with $T^3_f$ denoting the third component of the weak isospin of $f$ and
\begin{equation}\label{s_ast}
{s_*^2 = \hat s_{\mathrm{W}}^2 + \hat c_{\mathrm{W}}^2\frac{{{{\hat s}_{\mathrm{W}}}{t_\varepsilon }{t_\xi }}}{{1 + {{\hat s}_{\mathrm{W}}}{t_\varepsilon }{t_\xi }}}}.
\end{equation}
$j^\mu_\mathrm{DM} \propto g_X$ represents the $\mathrm{U}(1)_\mathrm{X}$ current of dark matter, which will be discussed in the following sections.
Such a current is coupled to $Z$ due to the kinetic mixing.
Furthermore, the neutral current coupled to $Z'$ can be expressed as
\begin{equation}\label{j_Zprime_mu}
j_{Z'}^\mu  = \frac{e({{\hat s}_{\mathrm{W}}}{t_\varepsilon }{c_\xi } - {s_\xi })}{{2{{\hat s}_{\mathrm{W}}}{{\hat c}_{\mathrm{W}}}}}\sum\limits_f {\bar f{\gamma ^\mu }(T_f^3 - 2{Q_f}\hat s_{\mathrm{W}}^2 - T_f^3{\gamma _5})f}
- {{\hat c}_{\mathrm{W}}}{t_\varepsilon }{c_\xi }j_{{\mathrm{EM}}}^\mu
+ \frac{{{c_\xi }}}{{{c_\varepsilon }}} j^\mu_\mathrm{DM}.
\end{equation}

Note that the photon couplings to SM fermions at tree level remain the same forms as in the SM.
The electroweak gauge couplings $\hat{g}$ and $\hat{g}'$ are related to the electric charge unit $e$ through $\hat{g}=e/\hat{s}_\mathrm{W}$ and $\hat{g}'=e/\hat{c}_\mathrm{W}$, where $e=\sqrt{4 \pi \alpha}$ can be determined by the $\overline{\mathrm{MS}}$ fine structure constant $\alpha(m_Z) = 1/127.955$ at the $Z$ pole~\cite{Tanabashi:2018oca}.

In the SM, the weak mixing angle satisfies
\begin{eqnarray}\label{sWcW}
s_\mathrm{W}^2c_\mathrm{W}^2 = \frac{{\pi \alpha }}{{\sqrt 2 {G_\mathrm{F}}m_Z^2}}
\end{eqnarray}
at tree level.
Based on this relation, one can define a ``physical" weak mixing angle $\theta_\mathrm{W}$ via the best measured parameters $\alpha$, $G_\mathrm{F}$, and $m_Z$~\cite{Burgess:1993vc,Babu:1997st}.
In the hidden $\mathrm{U}(1)_\mathrm{X}$ gauge theory, nonetheless, we have a similar relation,
\begin{equation}
\hat{s}_\mathrm{W}^2 \hat{c}_\mathrm{W}^2 = \frac{{\pi \alpha }}{{\sqrt 2 {G_\mathrm{F}} \hat{m}_Z^2}}.
\end{equation}
Therefore, the hatted weak mixing angle $\hat\theta_\mathrm{W}$ is related to $\theta_\mathrm{W}$ through $\hat{s}_\mathrm{W} \hat{c}_\mathrm{W} \hat{m}_{Z} = s_\mathrm{W} c_\mathrm{W} m_{Z}$.
Making use of Eq.~\eqref{massofgauge:Z}, we arrive at~\cite{Chun:2010ve}
\begin{equation}\label{weak_mixing_angle}
s_\mathrm{W}^2c_\mathrm{W}^2
= \frac{\hat{s}_\mathrm{W}^2\hat{c}_\mathrm{W}^2}{1+\hat{s}_\mathrm{W} t_{\varepsilon} t_{\xi}}.
\end{equation}

Hereafter, we adopt a free parameter set,
\begin{equation}
\{g_X,~ m_{Z'},~ m_s,~  s_{\varepsilon},~ s_{\eta}\}.
\end{equation}
From these free parameters, we can derive other parameters based on the above expressions.
As a result, both $\hat{s}_\mathrm{W}$ and $t_{\xi}$ become functions of $s_\varepsilon$ and $m_{Z'}$.
The relations between the free and induced parameters are further described in Appendix~\ref{param_relation}.
Current Higgs signal strength measurements at the LHC have given a constraint on the scalar mixing angle $\eta$ as $|s_\eta| \lesssim 0.37$ at 95\% confidence level (C.L.)~\cite{Ilnicka:2018def}.
We will choose appropriate values for $s_\eta$ in the following numerical analyses.

\subsection{Constraint from electroweak oblique parameters}

Because of the kinetic mixing, the electroweak oblique parameters $S$ and $T$~\cite{Peskin:1990zt,Peskin:1991sw} are modified at tree level.
Therefore, electroweak precision measurements have put a significant constraint on the kinetic mixing parameter $s_\varepsilon$.
Details of related electroweak precision tests can be found in Refs.~\cite{Burgess:1993vc,Babu:1997st,Chang:2006fp,Feldman:2007wj,Chun:2010ve,Frandsen:2011cg}.

In the effective Lagrangian formulation of the electroweak oblique parameters, the $Zff$ neutral current interactions can be expressed as~\cite{Burgess:1993vc}
\begin{equation}
\mathcal{L}_{Zff} = \frac{e}{{2{s_{\mathrm{W}}}{c_{\mathrm{W}}}}}\left( {1 + \frac{{\alpha T}}{2}} \right){Z_\mu }\sum\limits_f {\bar f{\gamma ^\mu }(T_f^3 - 2{Q_f}s_*^2 - T_f^3{\gamma _5})f},
\end{equation}
with
\begin{equation}
s_*^2 = s_{\mathrm{W}}^2 + \frac{1}{{c_{\mathrm{W}}^2 - s_{\mathrm{W}}^2}}\left( {\frac{{\alpha S}}{4} - s_{\mathrm{W}}^2c_{\mathrm{W}}^2\alpha T} \right).
\end{equation}
Comparing to Eqs.~\eqref{j_Z_mu}, \eqref{s_ast}, and \eqref{weak_mixing_angle}, we find that
\begin{eqnarray}\label{ST}
\alpha T &=& 2 c_\xi\sqrt{1+\hat{s}_\mathrm{W}t_\varepsilon t_\xi} - 2,\\
\alpha S&=&4 (c_\mathrm{W}^{2}-s_\mathrm{W}^{2}) \left(\hat{s}_\mathrm{W}^{2} -s_\mathrm{W}^{2} +\hat{c}_\mathrm{W}^{2} \frac{\hat{s}_\mathrm{W} t_{\varepsilon} t_{\xi}}{1+\hat{s}_\mathrm{W} t_{\varepsilon} t_{\xi}}\right)
+ 4 s_\mathrm{W}^{2} c_\mathrm{W}^{2} \alpha T.
\end{eqnarray}
Utilizing these expressions, we obtain $S$ and $T$ as functions of $s_\varepsilon$ and $m_{Z'}$.

\begin{figure}[!t]
	\centering
	\subfigure[~~$m_{Z'}<m_Z$\label{fig:oblique:small}]
	{\includegraphics[width=0.49\textwidth]{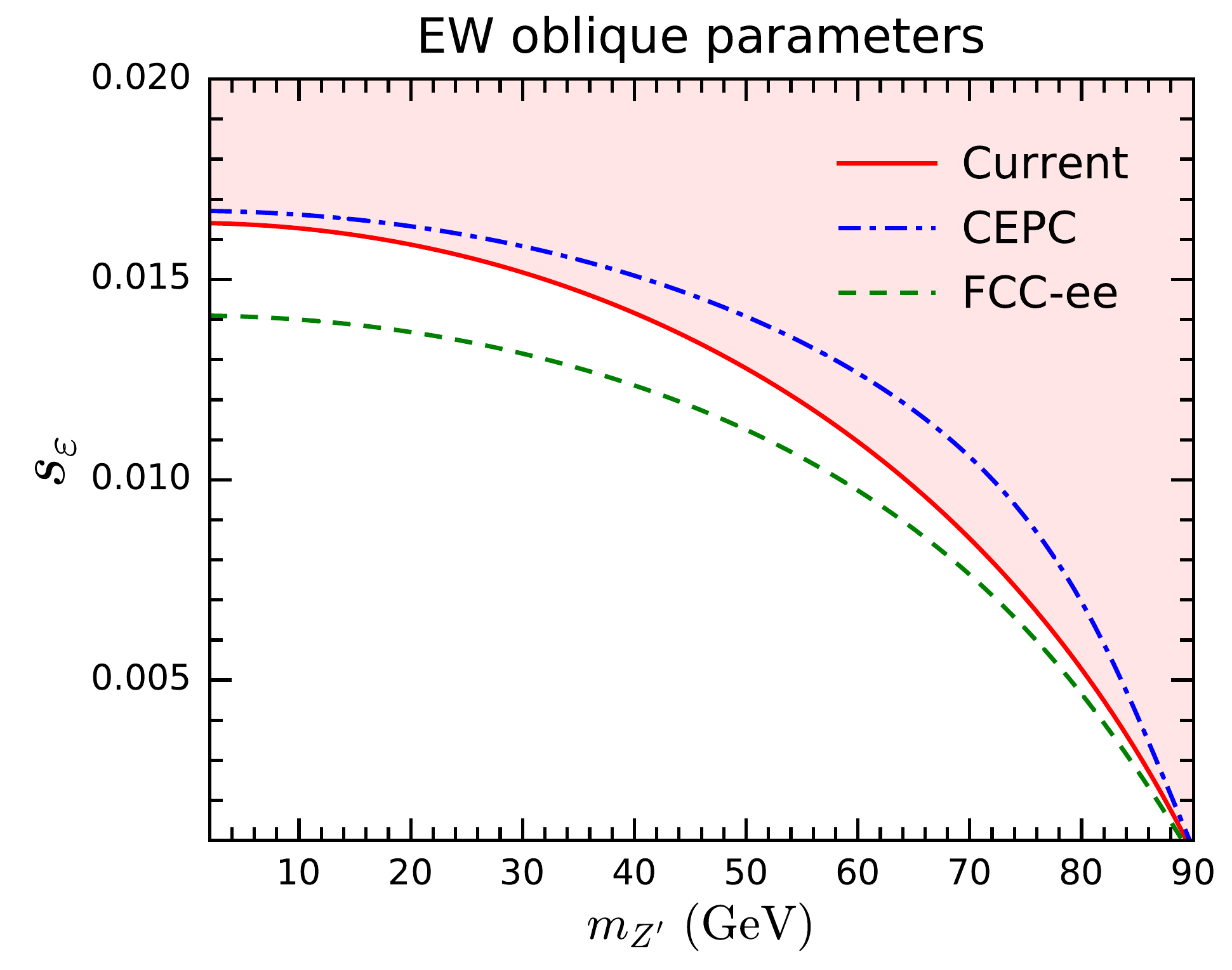}}
	\subfigure[~~$m_{Z'}>m_Z$\label{fig:oblique:large}]
	{\includegraphics[width=0.48\textwidth]{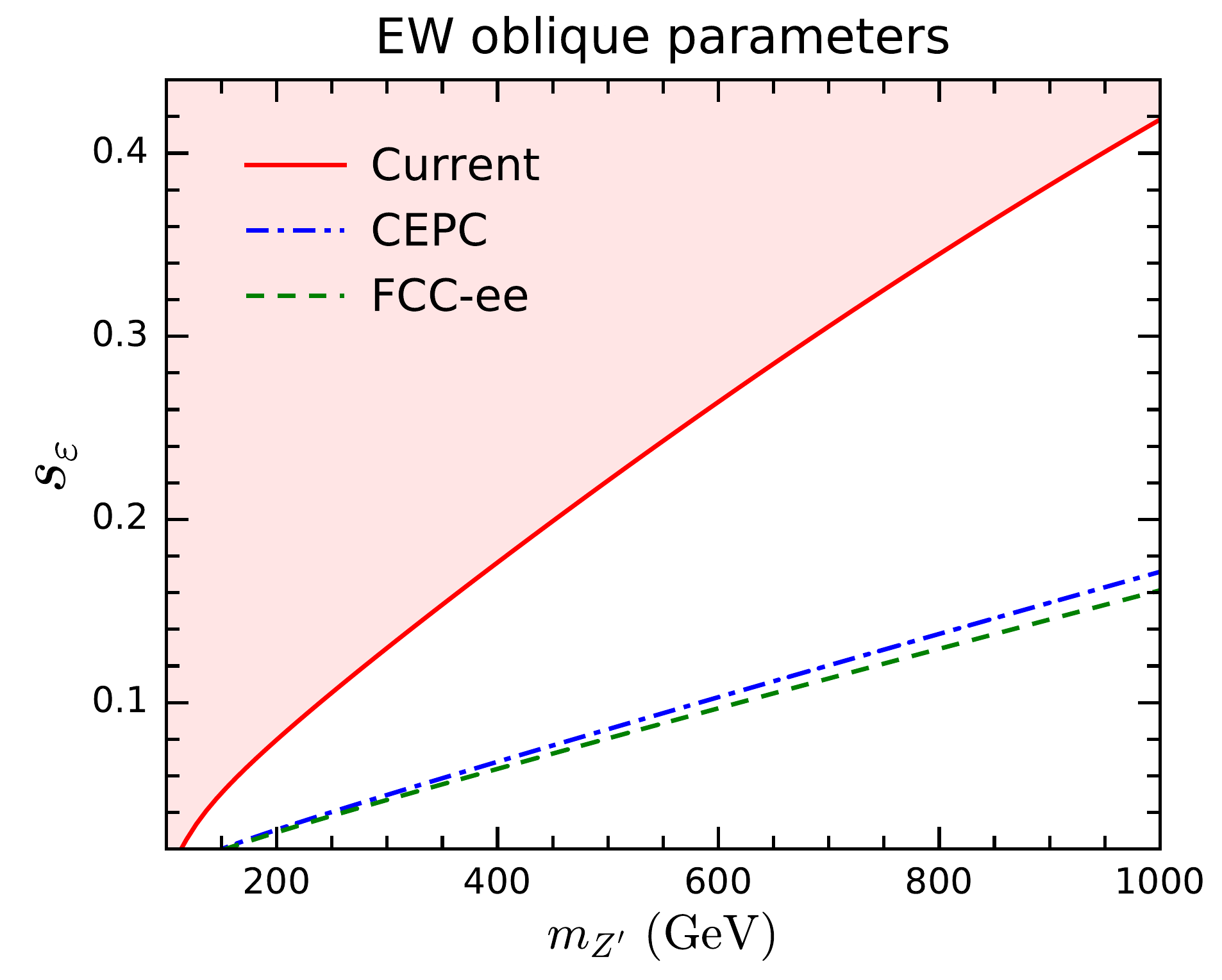}}
	\caption{95\% C.L. upper limits on the kinetic mixing  parameter $s_\varepsilon$ from the measurement of electroweak oblique parameters for the cases of $m_{Z'}<m_Z$ (a) and $m_{Z'}>m_Z$ (b).
    The red shaded regions are excluded by the global fit of current electroweak precision data from the Gfitter Group~\cite{Baak:2014ora}.
    The dot-dashed blue and dashed green lines correspond to the sensitivities in the future CEPC~\cite{CEPCStudyGroup:2018ghi} and FCC-ee~\cite{Abada:2019zxq} projects, respectively.}
	\label{fig:oblique}
\end{figure}

\begin{figure}[!t]
\centering
\includegraphics[width=0.5\textwidth]{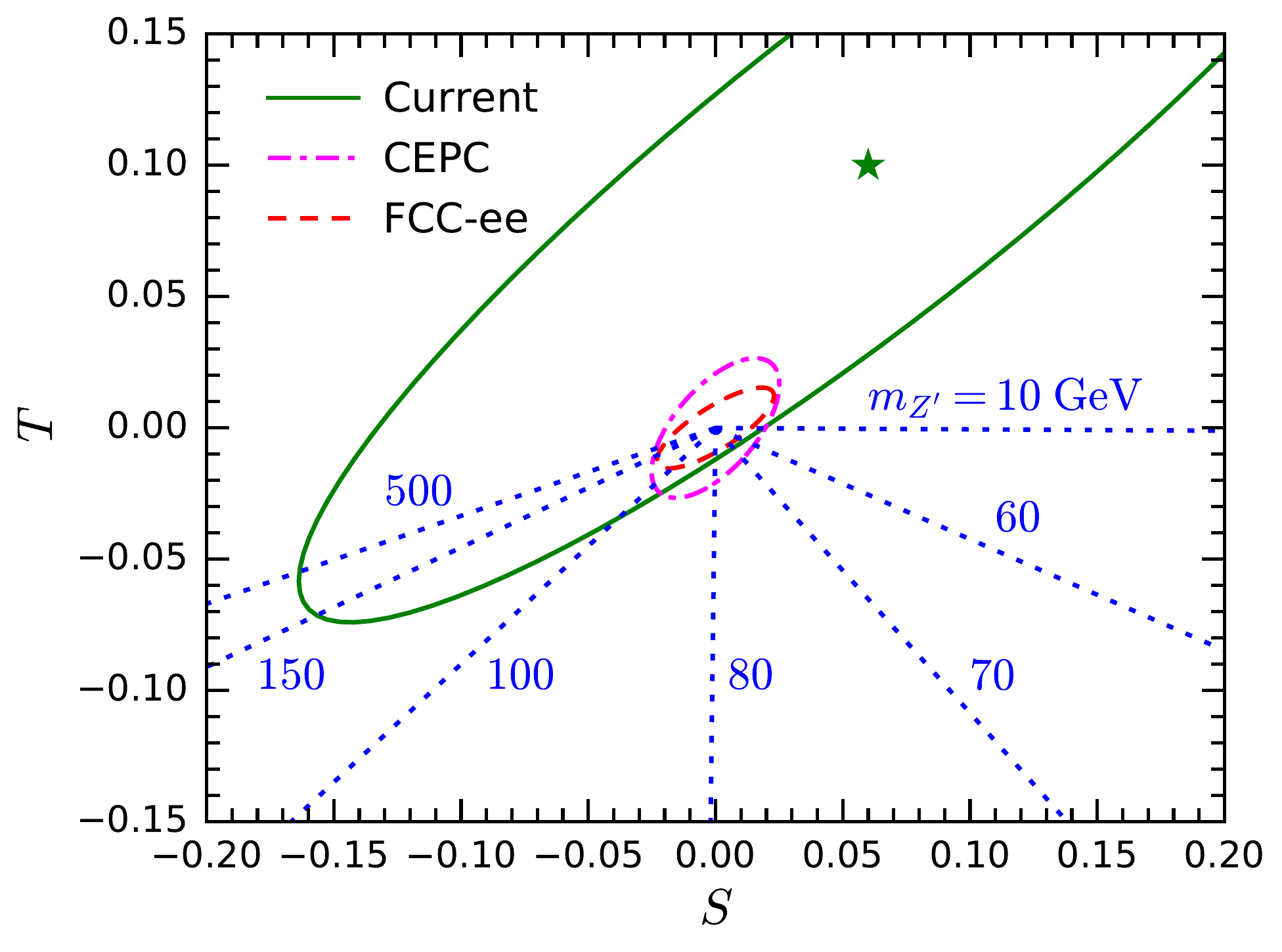}
\caption{Prediction of $S$ and $T$ for fixed $m_{Z'} = 10, 60, 70, 80, 100, 150, 500~\si{GeV}$ with varying $s_\varepsilon$.
The green curve denotes the current constraint at 95\% C.L. from the global fit of the Gfitter Group, while the corresponding central values are indicated by the green star.
The dot-dashed magenta and dashed red ellipses denote the projected 95\% C.L. sensitivity in the CEPC and FCC-ee experiments, respectively.}
\label{fig:ST_mzp}
\end{figure}

Assuming $U=0$, a global fit of electroweak precision data from the Gfitter Group gives~\cite{Baak:2014ora}
\begin{eqnarray}
S=0.06 \pm 0.09, \quad T=0.10 \pm 0.07,
\end{eqnarray}
with a correlation coefficient $\rho_{ST} = 0.91$.
Using this result, we derive upper limits on $s_\varepsilon$ at 95\% C.L., as shown in Fig.~\ref{fig:oblique}.
For a light $Z'$ ($r \ll 1$), $s_\varepsilon$ is bounded by $s_{\varepsilon} \lesssim 0.0165$.
For $m_{Z'}\sim 1~\si{TeV}$, the upper limit increases to $s_{\varepsilon} \sim 0.42$.
For $\varepsilon \ll 1$, $S$ and $T$ can be approximated as
\begin{eqnarray}
S \simeq \frac{{4s_{\mathrm{W}}^2c_{\mathrm{W}}^2{\varepsilon ^2}}}{\alpha{(1 - r)}}\left( {1 - \frac{{s_{\mathrm{W}}^2}}{{1 - r}}} \right),\quad
T \simeq - \frac{{rs_{\mathrm{W}}^2{\varepsilon ^2}}}{{\alpha{{(1 - r)}^2}}}.
\end{eqnarray}
When $r\sim 1$, the $(1-r)$ factors in the denominators greatly enlarge $|S|$ and $|T|$.
Therefore, the upper bound on $s_\varepsilon$ significantly decreases as $m_{Z'}$ closes to $m_Z$.
Moreover, these expressions mean that the ratio $T/S$ is basically independently of $s_\varepsilon$, and there is a linear relation between $S$ and $T$ for fixed $m_{Z'}$.
Such a linear relation is clearly shown by the dotted blue lines in Fig.~\ref{fig:ST_mzp} for fixed $m_{Z'}$ with varying $s_\varepsilon$.

Note that the current electroweak fit leads to central values $(S,T)=(0.06,0.10)$, and the SM prediction $(S,T)=(0,0)$ is quite close to the edge of the 95\% confidence ellipse, as demonstrated in Fig.~\ref{fig:ST_mzp}.
For $m_{Z'} \lesssim 80~\si{GeV}$, the kinetic mixing pushes $S$ and $T$ going through rather short paths out of the ellipse, leading to stringent constraints on $s_\varepsilon$.
On the other hand, $m_{Z'} \gtrsim 100~\si{GeV}$ leads to longer paths, and constraints on $s_\varepsilon$ are less stringent.

Future lepton collider projects, such as the Circular Electron-Positron Collider (CEPC)~\cite{CEPCStudyGroup:2018ghi} and the $e^+ e^-$ Future Circular Collider (FCC-ee)~\cite{Abada:2019zxq}, would significantly improve the precision of electroweak oblique parameters through measurements at the $Z$ pole and in the $W^+W^-$ threshold scan. 
According to the conceptual design report of CEPC~\cite{CEPCStudyGroup:2018ghi}, the projected precision of $S$ and $T$ measurements can be expressed as
\begin{equation}
\sigma_S = 0.0101,\quad
\sigma_T = 0.0107,\quad
\rho_{ST} = 0.624,
\end{equation}
with $\sigma_S$ and $\sigma_T$ denoting the $1\sigma$ uncertainties of $S$ and $T$.
Since FCC-ee could perform an additional $t\bar{t}$ threshold scan, its projected precision would be better than CEPC and reads~\cite{Fan:2014vta,Fedderke:2015txa}
\begin{equation}
\sigma_S = 0.00924,\quad
\sigma_T = 0.00618,\quad
\rho_{ST} = 0.794.
\end{equation}

As we have no information about the central values of $S$ and $T$ derived from future measurements, it is reasonable to use the SM prediction $(S,T) = (0,0)$ as the central values when evaluating the projected sensitivity to new physics~\cite{Cai:2016sjz,Cai:2017wdu}.
In this context, the projected 95\% C.L. sensitivities of CEPC and FCC-ee are presented as dot-dashed magenta and dashed red ellipses in Fig.~\ref{fig:ST_mzp}.
Although the CEPC precision is obviously much higher than current measurements, setting $(S,T) = (0,0)$ as the central values makes a fraction of the CEPC ellipse outside the current ellipse.
Therefore, the expected constraint on $s_\varepsilon$ from CEPC looks even weaker than the current one in the case of $m_{Z'} < m_Z$, as demonstrated in Fig.~\ref{fig:oblique:small}.
On the other hand, the expected FCC-ee constraint would be slightly stronger for $m_{Z'} < m_Z$.
In the case of $m_{Z'} > m_Z$ shown in Fig.~\ref{fig:oblique:large}, both CEPC and FCC-ee would be quite sensitive, reaching down to $s_\varepsilon \sim 0.16$ for $m_{Z'} = 1~\si{TeV}$.

\section{Dirac fermionic dark matter}
\label{sec:fermionic_DM}

In this section, we discuss a model where the DM particle is a Dirac fermion $\chi$ with $\mathrm{U}(1)_\mathrm{X}$ charge $q_\chi$~\cite{Mambrini:2010dq,Chun:2010ve,Liu:2017lpo,Bauer:2018egk}.
The Lagrangian for $\chi$ reads
\begin{eqnarray}
 \mathcal{L}_{\chi}=i \bar{\chi} \gamma^{\mu} D_{\mu} \chi-m_{\chi} \bar{\chi} \chi,
 \end{eqnarray}
where $D_\mu \chi  =(\partial_\mu - i q_\chi g_X \hat{Z}'_\mu)\chi$ and $m_\chi$ is the $\chi$ mass.
In this case, the DM neutral current appearing in Eqs.~\eqref{j_Z_mu} and \eqref{j_Zprime_mu} is
\begin{equation}
j^\mu_\mathrm{DM} = q_\chi g_X \bar\chi \gamma^\mu \chi.
\end{equation}
Thus, DM can communicate with SM fermions through the mediation of $Z$ and $Z'$ bosons, based on the kinetic mixing portal.
Through the thermal production mechanism, the number densities of $\chi$ and its antiparticle $\bar\chi$ should be equal, leading to a symmetric DM scenario.
Both $\chi$ and $\bar\chi$ particles constitute dark matter in the Universe.
Below we study the phenomenology in DM direct detection, as well as relic abundance and indirect detection.

\subsection{Direct detection}

In such a Dirac fermionic DM model, DM-quark interactions mediated by $Z$ and $Z'$ bosons could induce potential signals in direct detection experiments.
As DM particles around the Earth have velocities $\sim 10^{-3}$, these experiments essentially operate at zero momentum transfers.
In the zero momentum transfer limit, only the vector current interactions between $\chi$ and quarks contribute to DM scattering off nuclei in detectors.
Such interactions can be described by an effective Lagrangian (see, e.g., Ref.~\cite{Zheng:2010js}),
\begin{equation}\label{DM-q_scat}
\mathcal{L}_{\chi q}=\sum_{q} G^{\mathrm{V}}_{\chi q} \bar{\chi} \gamma^{\mu} \chi \bar{q} \gamma_{\mu} q,
\end{equation}
with $q=d,u,s,c,b,t$, and
\begin{equation}\label{G_V_chiq}
G^{\mathrm{V}}_{\chi q} = -\frac{q_{\chi} g_{X}}{c_{\varepsilon}}\left(\frac{s_{\xi} g_{Z}^{q}}{m_{Z}^{2}}+\frac{c_{\xi} g_{Z^{\prime}}^{q}}{m_{Z^{\prime}}^{2}}\right).
\end{equation}
From Eqs.~\eqref{j_Z_mu} and \eqref{j_Zprime_mu}, the vector current couplings of quarks to $Z$ and $Z'$ bosons can be expressed as
\begin{eqnarray}
g_{Z}^{q} &=& \frac{{e{c_\xi }}(1 + {{\hat s}_{\mathrm{W}}}{t_\varepsilon }{t_\xi })}{{2{{\hat s}_{\mathrm{W}}}{{\hat c}_{\mathrm{W}}}}}(T_q^3 - 2{Q_q}s_*^2),
\label{g_Z_q}\\
g_{Z^{\prime}}^{q} &=& \frac{e({{\hat s}_{\mathrm{W}}}{t_\varepsilon }{c_\xi } - {s_\xi })}{{2{{\hat s}_{\mathrm{W}}}{{\hat c}_{\mathrm{W}}}}}(T_q^3 - 2{Q_q}\hat s_{\mathrm{W}}^2) - {Q_q}e{{\hat c}_{\mathrm{W}}}{t_\varepsilon }{c_\xi }.
\label{g_Zprime_q}
\end{eqnarray}

The DM-quark interactions give rise to the DM-nucleon interactions, which can be described by an effective Lagrangian,
\begin{eqnarray}
\mathcal{L}_{\chi N}=\sum_{N =p,n} G^\mathrm{V}_{\chi N} \bar{\chi} \gamma^{\mu} \chi \bar{N} \gamma_{\mu} N,
\end{eqnarray}
where $N$ represents nucleons.
As the vector current counts the numbers of valence quarks in the nucleon, we have $G^\mathrm{V}_{\chi p}=2 G^\mathrm{V}_{\chi u}+G^\mathrm{V}_{\chi d}$
and $G^\mathrm{V}_{\chi n} = G^\mathrm{V}_{\chi u} + 2 G^\mathrm{V}_{\chi d}$.
Utilizing Eqs.~\eqref{G_V_chiq}, \eqref{g_Z_q}, \eqref{g_Zprime_q}, and \eqref{eq:t_xi}, we find that
\begin{equation}\label{cp_chi_N}
G^\mathrm{V}_{\chi p} = \frac{{{q_\chi }{g_X}e{{\hat c}_{\mathrm{W}}}{t_\varepsilon }c_\xi ^2(1 + t_\xi ^2r)}}{{{c_\varepsilon }m_{Z'}^2}},\quad
G^\mathrm{V}_{\chi n} = 0.
\end{equation}
The second expression means that $\chi n$ scattering vanishes in the zero momentum transfer limit.

\begin{figure}[!t]
	\centering
	\includegraphics[width=0.42\textwidth]{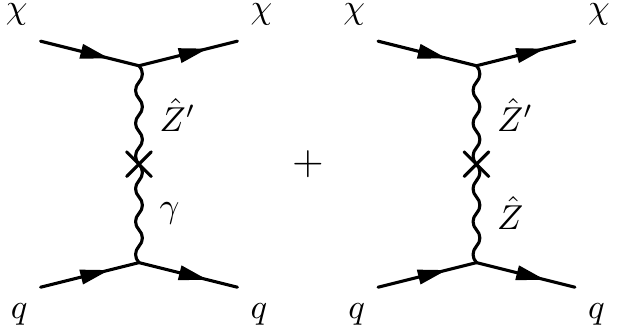}
	\caption{Feynman diagrams for $\chi q$ scattering. The crosses indicate the kinetic mixing term.}
	\label{fig:fd_chiq}
\end{figure}

A simple way to understand this is to realize that the kinetic mixing term $-s_\varepsilon \hat{B}^{\mu\nu}\hat{Z}'_{\mu\nu}/2$ contributes a $s_\varepsilon Q^2$ factor to the scattering amplitude, where $Q^\mu$ is the four-momentum of the mediator, i.e., the momentum transfer.
Note that the $\hat{B}^\mu$ field is related to the photon field $A^\mu$ by $\hat{B}^\mu = \hat{c}_\mathrm{W} A^\mu - \hat{s}_\mathrm{W} \hat{Z}^\mu$.
Thus, $\chi q$ scattering can be represented by two Feynman diagrams, as depicted in Fig.~\ref{fig:fd_chiq}.
In the zero momentum transfer limit, i.e., $Q^2 \to 0$, the $s_\varepsilon Q^2$ factor only picks up the $1/Q^2$ pole of the photon propagator in the first diagram,
while the second diagram vanishes because $\hat{Z}^\mu$ is massive.
Therefore, $\chi q$ scattering is essentially induced by the photon-mediated electromagnetic current $j_\mathrm{EM}^\mu$.
Since the neutron has no net electric charge, we arrive at $G^\mathrm{V}_{\chi n} = 0$, resulting in vanishing $\chi n$ scattering.

As ${G^\mathrm{V}_{\chi n}} = 0 \neq G^\mathrm{V}_{\chi p}$, isospin is violated in DM scattering off nucleons.
Thus, the conventional way for interpreting data in direct detection experiments, which assumes isospin conservation, is no longer suitable for our model.
Now we confront this issue following the strategy in Refs.~\cite{Feng:2011vu,Feng:2013fyw}.

For a nucleus $A$ constituted by $Z$ protons and $(A-Z)$ neutrons, the spin-independent (SI) $\chi A$ scattering cross section assuming a pointlike nucleus is
\begin{eqnarray}
\sigma_{\chi A}=\frac{\mu_{\chi A}^{2}}{\pi}\left[Z G^\mathrm{V}_{\chi p}+(A-Z) G^\mathrm{V}_{\chi n}\right]^{2},
\end{eqnarray}
where 
\begin{equation}
\mu_{\chi A} \equiv \frac{m_{\chi} m_{A}}{m_{\chi}+m_{A}}
\end{equation}
is the reduced mass of $\chi$ and $A$.
Note that the $\bar\chi A$ scattering cross section $\sigma_{\bar\chi A}$ is identical to $\sigma_{\chi A}$.
If isospin is conserved, i.e., $G^\mathrm{V}_{\chi p} = G^\mathrm{V}_{\chi n}$, we have $\sigma_{\chi A} =  A^2 \mu_{\chi A}^2 \sigma_{\chi p}/\mu_{\chi p}^2$, where
\begin{equation}
\sigma_{\chi p} = \frac{\mu_{\chi p}^{2} (G^\mathrm{V}_{\chi p})^{2}}{\pi}
\end{equation}
is the $\chi p$ scattering cross section with $\mu_{\chi p}$ denoting the reduced mass of $\chi$ and $p$.
Results in direct detection experiments are conventionally reported in terms of a normalized-to-nucleon cross section $\sigma_N^Z$ for SI scattering, assuming isospin conservation for detector material with an atomic number $Z$.
Therefore, in the isospin conservation case, we have $\sigma_N^Z = \sigma_{\chi p}$, and hence, a relation $\sigma_N^Z = \sigma_{\chi A}\mu_{\chi p}^2/(A^2 \mu_{\chi A}^2)$~\cite{Feng:2013fyw}.

Currently, the direct detection experiments utilizing two-phase xenon as detection material, including XENON1T~\cite{Aprile:2018dbl}, PandaX~\cite{Cui:2017nnn}, and LUX~\cite{Akerib:2016vxi}, are the most sensitive in the $5~\si{GeV} \lesssim m_\chi \lesssim 10~\si{TeV}$ range for SI scattering.
Among them, XENON1T gives the most stringent constraint.
Here, we would like to reinterpret its result for constraining our model.
Since xenon ($Z=54$) has several isotopes $A_i$, the event rate per unit time can be expressed as~\cite{Feng:2011vu}
\begin{equation}
R = {\sigma _{\chi p}}\sum\limits_i {{\eta _i}{I_{{A_i}}}\frac{{\mu _{\chi {A_i}}^2}}{{\mu _{\chi p}^2}}{{\left[ {Z + ({A_i} - Z)\frac{{{G^\mathrm{V}_{\chi n}}}}{{{G^\mathrm{V}_{\chi p}}}}} \right]}^2}},
\end{equation}
where $\eta_i$ is the fractional number abundance of $A_i$ in nature,
and $I_{A_i}$ is a factor depending on astrophysical, nuclear physics, and experimental inputs\footnote{The definition of $I_{A_i}$ can be found in Ref.~\cite{Feng:2011vu}.}.
For xenon, we have $A_i = \{128$, $129$, $130$, $131$, $132$, $134$, $136\}$, corresponding to $\eta_i = \{1.9\%$, $26\%$, $4.1\%$, $21\%$, $27\%$, $10\%$, $8.9\%\}$, respectively~\cite{Feng:2011vu}.

Experimentally, the normalized-to-nucleon cross section for SI scattering is determined in the isospin conservation case, where the relation $\sigma _N^Z = {\sigma _{\chi p}}$ holds.
This leads to
\begin{equation}\label{sigma_N_Z:IC}
\sigma _N^Z = \frac{R}{{\sum_i {{\eta _i}{I_{{A_i}}}A_i^2\mu _{\chi {A_i}}^2/\mu _{\chi p}^2} }}.
\end{equation}
In the isospin violation case, however, $\sigma _N^Z$ is not identical to $\sigma _{\chi p}$, which is given by
\begin{equation}
{\sigma _{\chi p}} = \frac{R}{{\sum_i {{\eta _i}{I_{{A_i}}}{{[Z + ({A_i} - Z){G^\mathrm{V}_{\chi n}}/{G^\mathrm{V}_{\chi p}}]}^2}\mu _{\chi {A_i}}^2/\mu _{\chi p}^2} }}.
\end{equation}
For a realistic situation, $I_{A_i}$ just varies mildly for different $A_i$, and thus, we can approximately assume that all $I_{A_i}$ are equal~\cite{Feng:2011vu}.
Therefore, the relation between $\sigma _N^Z$ and ${\sigma _{\chi p}}$ becomes
\begin{equation}
\sigma _N^Z = {\sigma _{\chi p}}\,\frac{{\sum_i {{\eta _i}\mu _{\chi {A_i}}^2{{[Z + ({A_i} - Z){G^\mathrm{V}_{\chi n}}/{G^\mathrm{V}_{\chi p}}]}^2}} }}{{\sum_i {{\eta _i}\mu _{\chi {A_i}}^2A_i^2} }}.
\end{equation}
This is the expression we should use when comparing the model prediction with the experimental results in terms of the normalized-to-nucleon cross section.

\begin{figure}[!t]
	\centering
	{\includegraphics[width=0.5\textwidth]{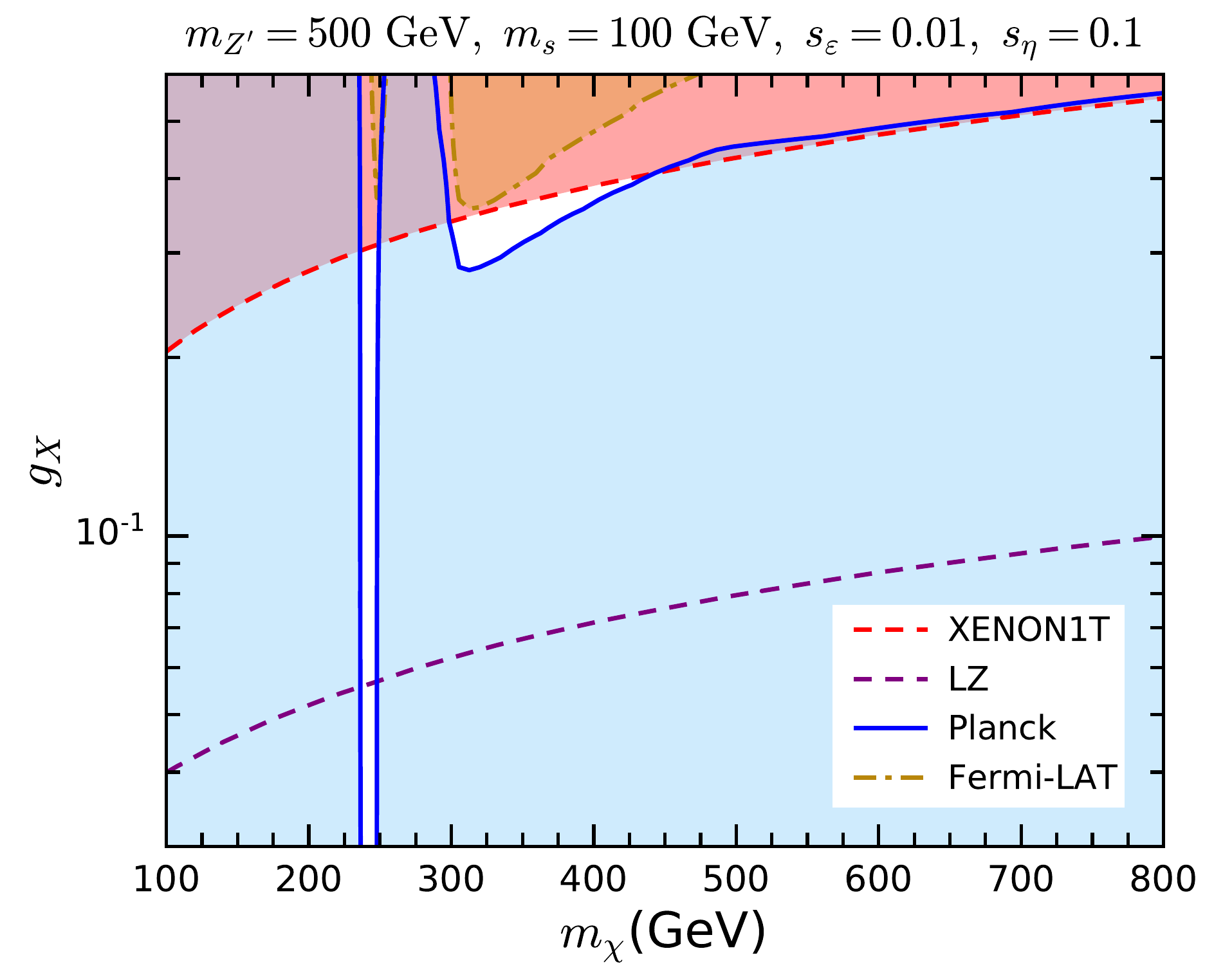}}
	\caption{Experimental constraints in the $m_\chi$-$g_X$ plane for Dirac fermionic DM with $m_{Z'} = 500~\si{GeV}$, $m_s = 100~\si{GeV}$, $s_\varepsilon = 0.01$, and $s_\eta = 0.1$.
    The red shaded area is excluded at 90\% C.L. by the XENON1T direct detection experiment~\cite{Aprile:2018dbl}.
    The dashed purple line denotes the 90\% C.L. sensitivity of the future LZ direct detection experiment~\cite{Mount:2017qzi}.
    The solid blue lines correspond to the mean value of the DM relic abundance, $\Omega_\mathrm{DM} h^2=0.120$, measured by the Planck experiment~\cite{Aghanim:2018eyx}, while the blue shaded areas indicate DM overproduction.
    The orange shaded areas are excluded at 95\% C.L. by the Fermi-LAT observations of dwarf galaxies~\cite{Ackermann:2015zua}.}
	\label{fig:mchi-gX:general}
\end{figure}

In our model, $G^\mathrm{V}_{\chi n}=0$, and the above expression reduces to
\begin{eqnarray}
\sigma _N^Z = \sigma_{\chi p}\,\frac{\sum_{i} \eta_{i} \mu_{\chi A_{i}}^{2} Z^{2}}{\sum_{i} \eta_{i} \mu_{\chi A_{i}}^{2} A_{i}^{2}}.
\end{eqnarray}
Therefore, $\sigma _N^Z$ is smaller than $\sigma_{\chi p}$, and experimental bounds are typically relaxed.
In the following numerical calculations, we adopt $q_\chi=1$ for simplicity.
Thus, $m_\chi$ is the only extra free parameter.
We use the 90\% C.L. upper bound on $\sigma _N^Z$ from the XENON1T experiment~\cite{Aprile:2018dbl} to obtain the exclusion region in the $m_\chi$-$g_X$ plane with fixed parameters $m_{Z'} = 500~\si{GeV}$, $m_s = 100~\si{GeV}$, $s_\varepsilon = 0.01$, and $s_\eta = 0.1$, as shown in Fig.~\ref{fig:mchi-gX:general}.
The $\mathrm{U}(1)_\mathrm{X}$ gauge coupling is constrained as $g_X \lesssim 0.2 \text{--} 0.55$ in the mass range $100~\textrm{GeV} \leq m_\chi \leq 800~\textrm{GeV}$.

Furthermore, we investigate the sensitivity of a future experiment LZ~\cite{Mount:2017qzi}, whose detection material is also two-phase xenon.
The corresponding expected exclusion limit at 90\% C.L. is demonstrated in Fig.~\ref{fig:mchi-gX:general}.
We find that LZ will be capable to reach down to $g_X \sim 0.04 \text{--} 0.1$ for $100~\textrm{GeV} \leq m_\chi \leq 800~\textrm{GeV}$.

\subsection{Relic abundance and indirect detection}

In the early Universe, $\chi$ and $\bar\chi$ particles would be produced in equal numbers via the thermal mechanism.
The total DM relic abundance is essentially determined by the total $\chi \bar\chi$ annihilation cross section at the freeze-out epoch.
The possible $\chi\bar\chi$ annihilation channels include $f\bar{f}$, $W^+W^-$, $h_i h_j$, $Z_iZ_j$, and $h_iZ_j$, with $h_i\in\{h,s\}$ and $Z_i \in \{Z,Z'\}$.
All these channels are mediated via $s$-channel $Z$ and $Z'$ bosons.
In addition, the $Z_iZ_j$ channels are also mediated via $t$- and $u$-channel $\chi$ propagators.

Some numerical tools are utilized to evaluation the prediction of the DM relic abundance in our model.
Firstly, we use a \textit{Mathematica} package  \texttt{FeynRules}~\cite{Alloul:2013bka} to generate model files, which encode the information of particles, Feynman rules, and parameter relations.
Then we interface the model files to a Monte Carlo generator \texttt{MadGraph5\_aMC@NLO}~\cite{Alwall:2014hca}.
Finally we invoke a \texttt{MadGraph} plugin \texttt{MadDM}~\cite{Backovic:2013dpa,Backovic:2015cra,Ambrogi:2018jqj} to calculate the relic abundance.
In the calculation, all possible annihilation channels are included, and the particle decay widths are automatically computed inside \texttt{MadGraph}.

From the measurement of cosmic microwave background anisotropies, the Planck experiment derives an observation value of the DM relic abundance, $\Omega_\mathrm{DM} h^2=0.120\pm0.001$~\cite{Aghanim:2018eyx}.
In Fig.~\ref{fig:mchi-gX:general}, the solid blue lines are corresponding to the mean value of $\Omega_\mathrm{DM} h^2$ predicted by the model.
In the blue shaded areas, the model predicts overproduction of dark matter, contradicting the cosmological observation.
On the other hand, a relic abundance lower than the observation value is not necessarily considered to be ruled out, as $\chi$ and $\bar\chi$ particles could only constitute a fraction of dark matter, or there could be extra nonthermal production of $\chi$ and $\bar\chi$ in the cosmological history.

In Fig.~\ref{fig:mchi-gX:general}, the kinetic mixing parameter we adopt, $s_\varepsilon = 0.01$, is rather small.
Thus, DM annihilation for $m_\chi \lesssim 230~\si{GeV}$ is commonly suppressed, leading to DM overproduction.
Nonetheless, the $Z'$-pole resonance effect at $m_\chi \sim m_{Z'}/2 = 250~\si{GeV}$ significantly enhances the annihilation cross section, giving rise to a narrow available region.
Moreover, the $sZ'$ and $Z'Z'$ annihilation channels opening for $m_\chi \gtrsim (m_s+m_{Z'})/2$ and $m_\chi \gtrsim m_{Z'}$ also greatly enhance the total annihilation cross section, because they are basically dark sector processes that are not suppressed by $s_\varepsilon$.
As a result, the solid blue curve with $m_\chi \gtrsim 280~\si{GeV}$ can give a correct relic abundance.

In addition, DM annihilation at present day could give rise to high energy $\gamma$ rays from the radiations and decays of the annihilation products.
Nonetheless, the Fermi-LAT experiment has reported no such signals in the continuous-spectrum observations of fifteen DM-dominated dwarf galaxies around the Milky Way with six-year data, leading to stringent bounds on the DM annihilation cross section~\cite{Ackermann:2015zua}.

We further utilize \texttt{MadDM} to calculate the total velocity-averaged DM annihilation cross section $\left<\sigma_\mathrm{ann} v\right>$ at a typical average velocity in dwarf galaxies, $2\times 10^{-5}$.
Then the Fermi-LAT 95\% C.L. upper limits on the annihilation cross section in the $b\bar{b}$ channel~\cite{Ackermann:2015zua} are adopted to constrain $\left<\sigma_\mathrm{ann} v\right>$.
This should be a good approximation, because the $\gamma$-ray spectra yielded from the dominant annihilation channels in our model would be analogue to that from the $b\bar{b}$ channel~\cite{Cirelli:2010xx}.
The orange shaded areas in Fig.~\ref{fig:mchi-gX:general} are excluded by the Fermi-LAT data.

In Fig.~\ref{fig:mchi-gX:general}, we can see that the relic abundance observation tends to disfavor small $g_X$, while the direct and indirect detection experiments tend to disfavor large $g_X$.
This leaves only two surviving regions.
One is a narrow strip around $m_\chi \sim m_{Z'}/2$ due to the $Z'$-pole resonance annihilation,
while the other region lies in $300~\si{GeV} \lesssim m_\chi \lesssim 450~\si{GeV}$, where the $sZ'$ annihilation channel opens. 

\begin{figure}[!t]
	\centering
	\subfigure[~~Fixed relation $m_{Z'}=2.05m_{\chi}$\label{fig:mchi-gX:relation:Zprime_pole}]
	{\includegraphics[height=0.4\textwidth]{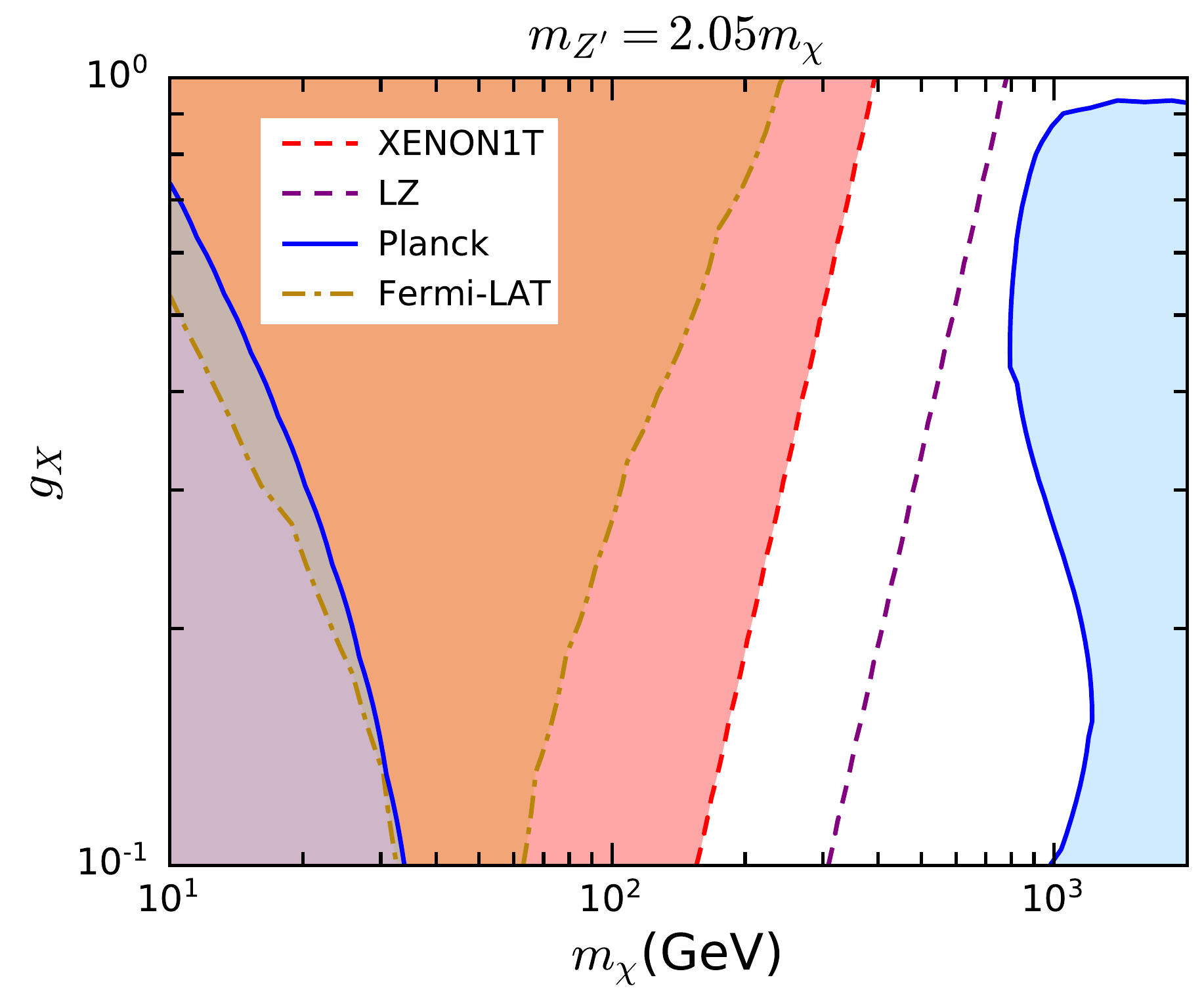}}
	\subfigure[~~Fixed relation $m_{Z'}=0.9(2m_{\chi}-m_s)$\label{fig:mchi-gX:relation:sZprime}]
	{\includegraphics[height=0.4\textwidth]{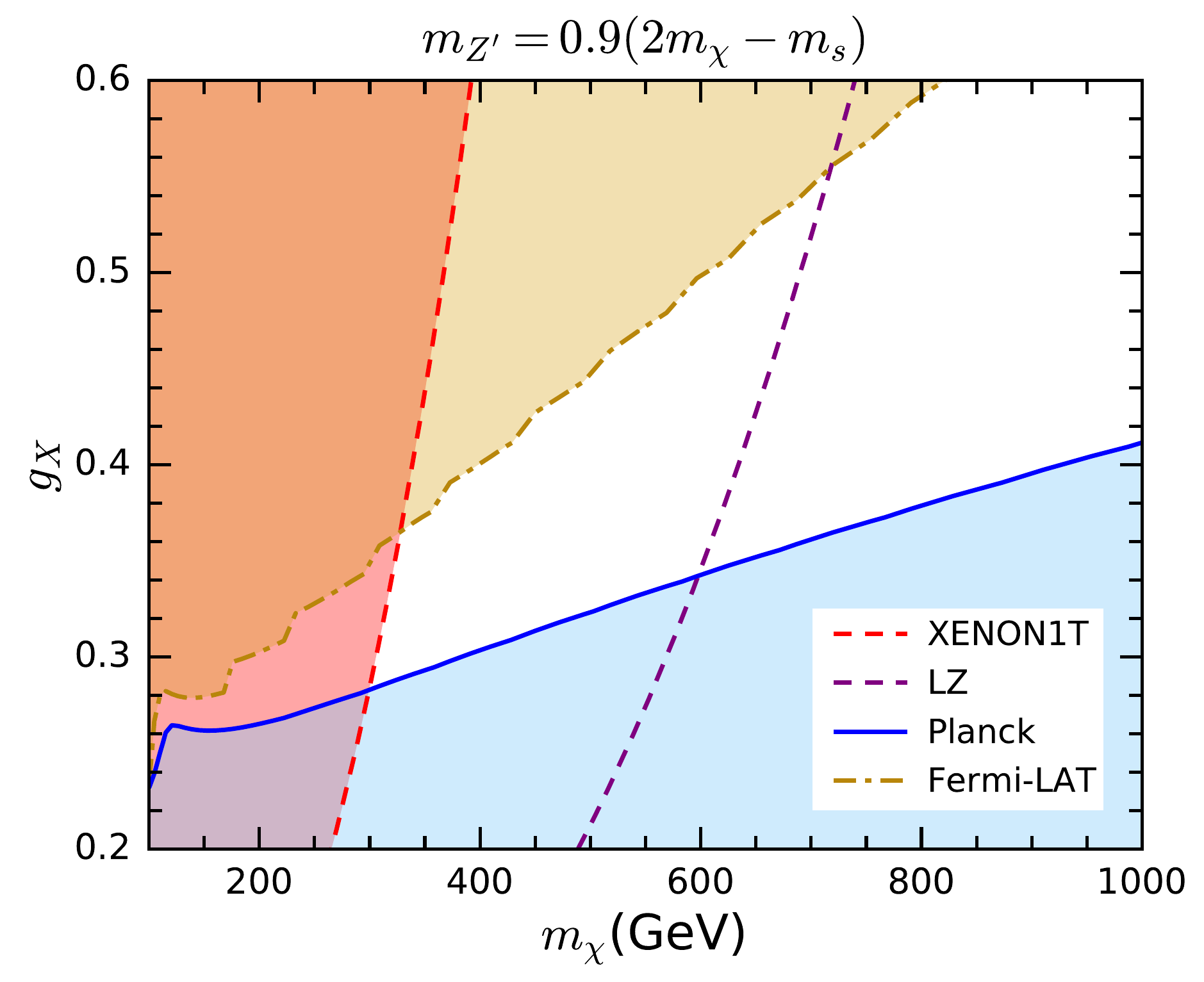}}
	\caption{Experimental constraints in the $m_\chi$-$g_X$ plane for Dirac fermionic DM with fixed relations $m_{Z'}=2.05m_{\chi}$ (a) and $m_{Z'}=0.9(2m_{\chi}-m_s)$ (b).
    The common parameters in both panels are $m_s=100~\si{GeV}$, $s_\varepsilon=0.01$, and $s_\eta=0.1$.}
	\label{fig:mchi-gX:relation}
\end{figure}

Now we explore more deeply into the parameter space.
Inspired by the above observation, we investigate the $Z'$ resonance region with a fixed relation $m_{Z'}=2.05m_\chi$ and demonstrate the result in Fig.~\ref{fig:mchi-gX:relation:Zprime_pole}.
Other parameters are chosen to be $m_s=100~\si{GeV}$, $s_\varepsilon=0.01$, and $s_\eta=0.1$.
We find that the correct relic abundance corresponds to two curves, one around $m_\chi \sim 10\text{--}30~\si{GeV}$ and one around $m_\chi \sim 1~\si{TeV}$.
A large area between the two curves predicts a lower relic abundance.
Nonetheless, the direct and indirect detection experiments have excluded a region with $m_\chi \lesssim 160\text{--}400~\si{GeV}$, which involves the first curve.
The second curve is totally allowed and beyond the probe of the LZ experiment.

Furthermore, we change the fixed relation to be $m_{Z'}=0.9(2m_{\chi}-m_s)$, with which the $sZ'$ annihilation channel always opens, and present the result in Fig.~\ref{fig:mchi-gX:relation:sZprime}.
The correct relic abundance is corresponding to a curve with $g_X\sim 0.23 \text{--} 0.41$ in the $100~\si{GeV} \leq m_\chi \leq 1~\si{TeV}$ range, which is not excluded by the Fermi-LAT data.
Nonetheless, the XENON1T experiment has excluded a region with $m_\chi \lesssim 270 \text{--} 400~\si{GeV}$, and the LZ experiment can explore up to $m_\chi \sim 740~\si{GeV}$.

\section{Complex Scalar Dark Matter}
\label{sec:scalar_DM}

For the Dirac fermionic DM model in the previous section, DM interactions with SM particles are only induced by the kinetic mixing portal.
Thus, the interaction strengths and types are limited.
As a result, it is not easy to simultaneously satisfy the direct detection and relic abundance constraints, except for some particular regions.
This motivates us to study complex scalar DM with an additional Higgs portal in this section. 

In the complex scalar DM model, we introduce a complex scalar field $\phi$ with $\mathrm{U}(1)_\mathrm{X}$ charge $q_\phi$.
The Lagrangian related to $\phi$ reads
\begin{eqnarray}\label{L_phi}
\mathcal{L}_{\phi}=\left(D^{\mu} \phi\right)^{\dagger}\left(D_{\mu} \phi\right)-\mu_{\phi}^{2} \phi^{\dagger} \phi
+\lambda_{S \phi} \hat{S}^{\dagger} \hat{S} \phi^{\dagger} \phi
+\lambda_{H \phi} \hat{H}^{\dagger} \hat{H} \phi^{\dagger} \phi
+ {\lambda _\phi }{({\phi ^\dag }\phi )^2},
\end{eqnarray}
where $D_\mu \phi = ( \partial_\mu - i q_\phi g_X \hat{Z}'_\mu)\phi$.
We assume that the $\phi$ field does not develop a VEV, and thus, the scalar boson $\phi$ and its antiparticle $\bar\phi$ are stable, serving as DM particles.
After $\hat{H}$ and $\hat{S}$ gain their VEVs, the mass squared of $\phi$ is given by
\begin{equation}
m_\phi ^2 = \mu _\phi ^2 - \frac{1}{2}{\lambda _{S\phi }}v_S^2 - \frac{1}{2}{\lambda _{H\phi }}{v^2}.
\end{equation}
The DM neutral current in Eqs.~\eqref{j_Z_mu} and \eqref{j_Zprime_mu} is
\begin{equation}
j^\mu_\mathrm{DM} = {q_\phi }{g_X} {\phi ^\dag }i\overleftrightarrow {{\partial ^\mu }}\phi,
\end{equation}
with ${\phi ^\dag }\overleftrightarrow {{\partial ^\mu }}\phi \equiv \phi^\dag \partial ^\mu \phi - (\partial ^\mu \phi^\dag) \phi$, leading to $\phi$ couplings to the $Z$ and $Z'$ bosons.
Besides, $\phi$ also couples to the scalar bosons $h$ and $s$, described by the Lagrangian,
\begin{equation}\label{eq:L_phi_h_s}
{\mathcal{L}_{\phi hs}} = ({\lambda _{S\phi }}{s_\eta }{v_S} + {\lambda _{H\phi }}{c_\eta }v)h{\phi ^\dag }\phi  + ({\lambda _{S\phi }}{c_\eta }{v_S} - {\lambda _{H\phi }}{s_\eta }v)s{\phi ^\dag }\phi .
\end{equation}

Note that for allowing the neutral current interactions between $\phi$ and SM fermions through the kinetic mixing portal, a global $\mathrm{U}(1)$ symmetry $\phi\to e^{i\theta} \phi$ should be preserved after the spontaneous symmetry breaking of $\mathrm{U}(1)_\mathrm{Y} \times \mathrm{U}(1)_\mathrm{X}$.
Such a global symmetry ensures $\phi$ being a complex scalar boson (i.e., the real and imaginary components of $\phi$ are degenerate in mass) and prevents $\phi$ from decaying.
Therefore, scalar interaction terms that violate this symmetry, such as $\hat{S}^\dag \hat{S}^\dag \hat{S}^\dag \phi$, $\hat{S}^\dag \hat{S}^\dag \phi$, $\hat{S}^\dag \hat{S}^\dag \phi \phi$, $\hat{S}^\dag \phi \phi$, $\hat{S}^\dag \phi \phi \phi$, and their Hermitian conjugates, should be forbidden from the beginning.
This can be achieved by assigning $q_\phi \neq \pm 3, \pm 2, \pm 1, \pm 1/2, \pm 1/3$.
Since there is no reason for the quantization of $\mathrm{U}(1)_\mathrm{X}$ charges, $q_\phi$ can be any real number except the above values.
For simplicity, we just fix $q_\phi = 1/4$ in the following numerical analyses, rather than treat it as a free parameter.

Now DM interactions with SM fermions are not only mediated by the $Z$ and $Z'$ bosons from the kinetic mixing portal, but also mediated by the $h$ and $s$ bosons as a Higgs portal.
Assuming $\phi$ and $\bar\phi$ particles are thermally produced in the early Universe, we arrive at a symmetric DM scenario; i.e., the present number densities of $\phi$ and $\bar\phi$ are equal.
However, as we will see soon, the $\phi A$ and $\bar\phi A$ scattering cross sections are not identical in general.

\subsection{Direct detection}

\begin{figure}[!t]
	\centering
    \subfigure[~~$\phi q$ scattering\label{fig:fd_phi-q:phi}]
    {\includegraphics[width=0.42\textwidth]{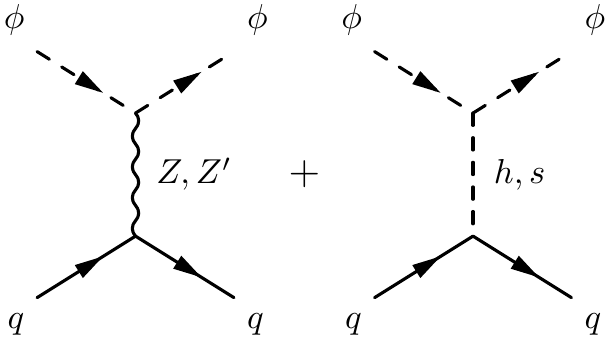}}
    \hspace{2em}
    \subfigure[~~$\bar\phi q$ scattering\label{fig:fd_phi-q:phibar}]
    {\includegraphics[width=0.42\textwidth]{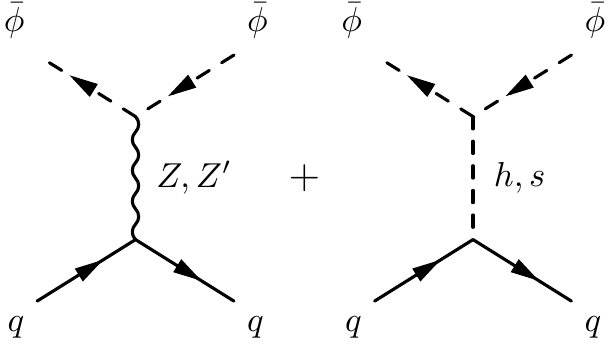}}
	\caption{Feynman diagrams for $\phi q$ (a) and $\bar\phi q$ (b) scatterings.}
	\label{fig:fd_phi-q}
\end{figure}

$\phi q$ and $\bar\phi q$ scatterings, which are relevant to direct detection, are mediated by the $Z$ and $Z'$ vector bosons (kinetic mixing portal) as well as by the $h$ and $s$ scalar bosons (Higgs portal).
The corresponding Feynman diagrams are depicted in Fig.~\ref{fig:fd_phi-q}.
In the zero momentum transfer limit, DM-quark interactions can be described by an effective Lagrangian  (see, e.g., Ref.~\cite{Yu:2011by}),
\begin{equation}
\mathcal{L}_{\phi q} = \sum_{q}\left[G^\mathrm{V}_{\phi q}(\phi^{\dagger} i \overleftrightarrow{\partial^{\mu}} \phi) \bar{q} \gamma_{\mu} q+G^\mathrm{S}_{\phi q} \phi^\dag \phi \bar{q} q\right].
\end{equation}
Similar to Eq.~\eqref{G_V_chiq}, the vector current effective coupling due to the kinetic mixing portal is
\begin{equation}\label{form_scalar}
G^\mathrm{V}_{\phi q} = -\frac{q_{\phi} g_{X} }{c_{\varepsilon}}\left(\frac{s_{\xi}g_{Z}^{q}}{m_{Z}^{2}} +\frac{c_{\xi}g_{Z^{\prime}}^{q}}{m_{Z^{\prime}}^{2}} \right),
\end{equation}
with $g_{Z}^{q}$ and $g_{Z^{\prime}}^{q}$ defined in Eqs.~\eqref{g_Z_q} and \eqref{g_Zprime_q}.
The scalar-type effective coupling induced by the Higgs portal is
\begin{equation}
G^\mathrm{S}_{\phi q} = \frac{m_{q}}{v}\left[\frac{s_{\eta}}{m_{s}^{2}}\left(\lambda_{S \phi} c_{\eta} v_{S} - \lambda_{H \phi} s_{\eta} v\right)-\frac{c_{\eta}}{m_{h}^{2}}\left(\lambda_{S \phi} s_{\eta} v_{S}+\lambda_{H \phi} c_{\eta} v\right)\right].
\end{equation}

At the nucleon level, the effective Lagrangian reads
\begin{equation}\label{L_phiN}
{\mathcal{L}_{\phi N}} = \sum\limits_{N = p,n} {\left[{G^{\mathrm{V}}_{\phi N}}({\phi ^\dag }i\overleftrightarrow {{\partial ^\mu }}\phi )\bar N{\gamma _\mu }N
+{G^{\mathrm{S}}_{\phi N}}{\phi ^\dag }\phi \bar NN \right]}.
\end{equation}
Analogous to the Dirac fermionic DM case, the vector current effective couplings for the proton and neutron are 
${G^\mathrm{V}_{\phi p}} = 2{G^\mathrm{V}_{\phi u}} + {G^\mathrm{V}_{\phi d}}$ and ${G^\mathrm{V}_{\phi n}} = {G^\mathrm{V}_{\phi u}} + 2{G^\mathrm{V}_{\phi d}}$.
Similar to Eqs.~\eqref{cp_chi_N}, we have
\begin{equation}\label{cp_phi_N}
G^\mathrm{V}_{\phi p} = \frac{{{q_\phi }{g_X}e{{\hat c}_{\mathrm{W}}}{t_\varepsilon }c_\xi ^2(1 + t_\xi ^2r)}}{{{c_\varepsilon }m_{Z'}^2}},\quad
G^\mathrm{V}_{\phi n} = 0.
\end{equation}
Once again, $G^\mathrm{V}_{\phi n}$ vanishes because the neutron does not carry electric charge.
On the other hand, the scalar-type effective couplings for nucleons are given by~\cite{Jungman:1995df}
\begin{eqnarray}
G^\mathrm{S}_{\phi N} = m_N \sum_{q=d, u, s} \frac{G^\mathrm{S}_{\phi q} f_{q}^{N}}{m_{q}}
+ m_N f_{Q}^{N} \sum_{q=c, b, t} \frac{G^\mathrm{S}_{\phi q}}{m_{q}} .
\end{eqnarray}
The form factors $f^N_q$ in the first term are related to light quark contributions to the nucleon mass, defined by ${m_N} f_q^N = \langle N |{m_q}\bar q q|N \rangle$.
Their values are $f_{u}^{p}=0.020 \pm 0.004$, $f_{d}^{p}=0.026 \pm 0.005$, $f_{u}^{n}=0.014 \pm 0.003$, $f_{d}^{n}=0.036 \pm 0.008$, $f_{s}^{p} = f_{s}^{n} = 0.118 \pm 0.062$~\cite{Ellis:2000ds}.
The second term with the form factor $f_Q^N = 2({1- f_d^N - f_u^N - f_s^N})/27$ is contributed by the heavy quarks at loop level.
An approximate relation $G^\mathrm{S}_{\phi p} \simeq G^\mathrm{S}_{\phi n}$ numerically holds~\cite{Belanger:2013tla}.
This means that the scalar-type interactions are roughly isospin conserving.

The $\phi N$ and $\bar\phi N$ scattering cross sections due to the Lagrangian \eqref{L_phiN} are obtained as
\begin{equation}
\sigma_{\phi N} = \frac{\mu_{\phi N}^{2} f_{\phi N}^{2}}{\pi},\quad
\sigma_{\bar\phi N} = \frac{\mu_{\phi N}^{2} f_{\bar\phi N}^{2}}{\pi},
\end{equation}
with
\begin{equation}
{{f_{\phi N}} = \frac{{{G^{\mathrm{S}}_{\phi N}}}}{{2{m_\phi }}} + {G^{\mathrm{V}}_{\phi N}}},\quad
{{f_{\bar \phi N}} = \frac{{{G^{\mathrm{S}}_{\phi N}}}}{{2{m_\phi }}} - {G^{\mathrm{V}}_{\phi N}}}.
\end{equation}
The only difference between the Feynman diagrams for the $\phi q$ and $\bar\phi q$ scatterings in Fig.~\ref{fig:fd_phi-q} is the arrow direction of the $\phi$ line, which affects the relative signs between the contributions from the vector current and scalar-type interactions.
This explains the different signs in the above $f_{\phi N}$ and $f_{\bar \phi N}$ expressions~\cite{Belanger:2008sj,Belanger:2013tla}.
Since $G^\mathrm{V}_{\phi n} = 0$, we have $f_{\phi n} = f_{\bar\phi n} = G^\mathrm{S}_{\phi n}/(2m_\phi)$.

\begin{figure}[!t]
\centering
\includegraphics[width=0.5\textwidth]{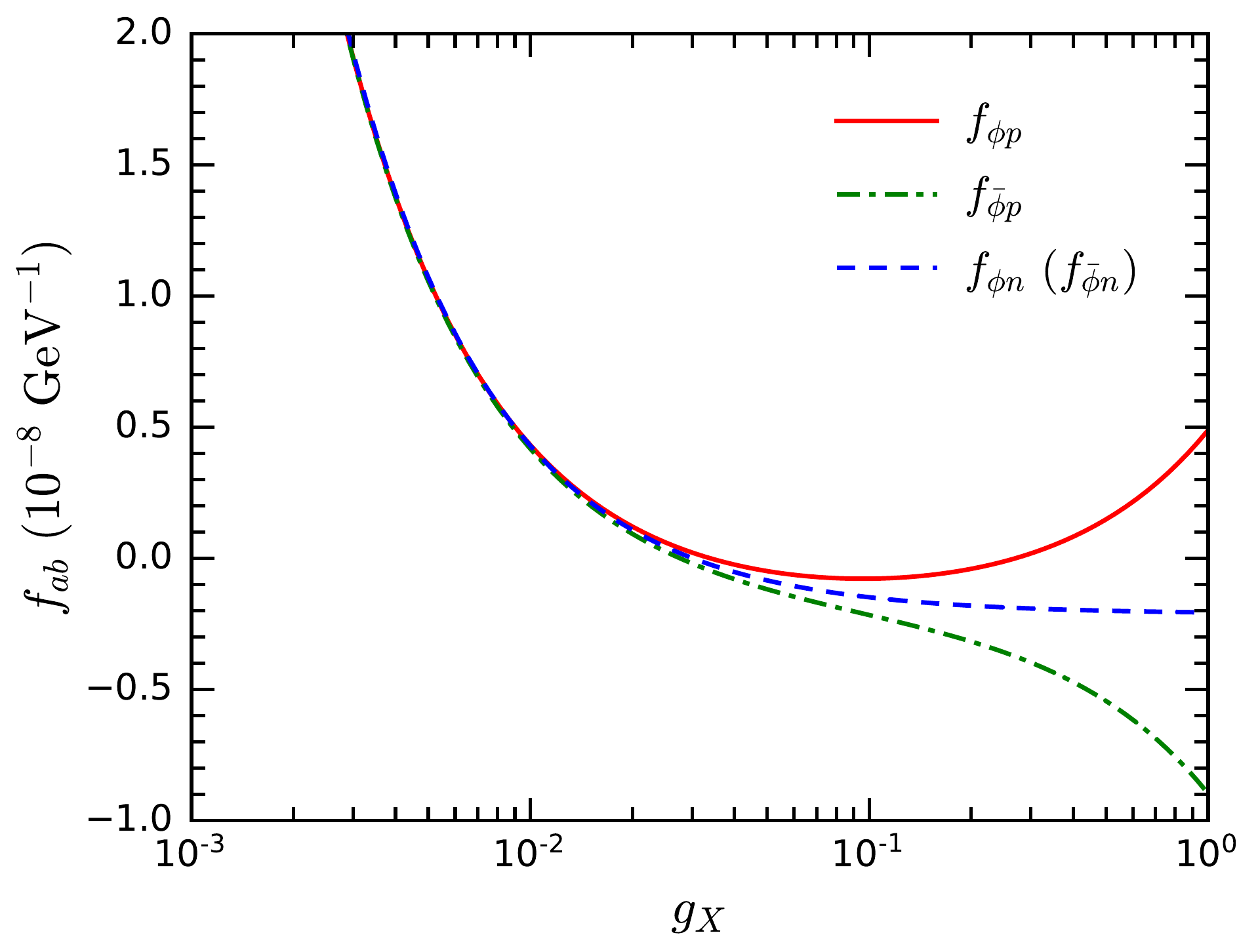}
\caption{$f_{\phi p}$, $f_{\bar\phi p}$, and $f_{\phi n}$ as functions of $g_X$ with $m_\phi=500~\si{GeV}$, $m_{Z'}=1~\si{TeV}$, $m_s=250~\si{GeV}$, $s_\varepsilon=0.1$, $s_\eta=0.01$, $\lambda_{H\phi }=0.1$, and $\lambda_{S\phi }=-0.1$.
Note that $f_{\bar\phi n} = f_{\phi n} = G^\mathrm{S}_{\phi n}/(2m_\phi)$.}
\label{fig:f_phiN}
\end{figure}

In Fig.~\ref{fig:f_phiN}, we demonstrate $f_{\phi p}$, $f_{\bar\phi p}$, and $f_{\phi n}$ as functions of $g_X$ for the fixed parameters $m_\phi=500~\si{GeV}$, $m_{Z'}=1000~\si{GeV}$, $m_s=250~\si{GeV}$, $s_\varepsilon=0.1$, $s_\eta=0.01$, $\lambda_{H\phi }=0.1$, and $\lambda_{S\phi }=-0.1$.
For $g_X \lesssim 0.03$, $f_{\phi p}$, $f_{\bar\phi p}$, and $f_{\phi n}$ are rather close to each other.
The reason is that the relation $G^\mathrm{S}_{\phi p} \simeq G^\mathrm{S}_{\phi n}$ holds and the contributions from $G^\mathrm{V}_{\phi p}$ are negligible for small $g_X$.
From Eq.~\eqref{v_S}, we know that $v_S \propto 1/g_X$.
Consequently, as $g_X$ increases, $G^\mathrm{S}_{\phi p}$ and $G^\mathrm{S}_{\phi n}$ decrease,
and hence, $f_{\phi p}$, $f_{\bar\phi p}$, and $f_{\phi n}$ decrease till $g_X \sim 0.03$, where they close to zero.
At $g_X \sim 0.03$, the contributions from the $h$ and $s$ mediators roughly cancel each other out, and thus, $G^\mathrm{S}_{\phi p}$ and $G^\mathrm{S}_{\phi n}$ basically vanish.
After this point, the contributions from $G^\mathrm{V}_{\phi p}$ become more and more important, pushing $f_{\phi p}$ up but lowering $f_{\bar\phi p}$ down.

Note that $f_{\phi n} = f_{\bar\phi n}$ leads to $\sigma_{\phi n} = \sigma_{\bar\phi n}$.
Nonetheless, $\sigma_{\phi p}$ and $\sigma_{\bar\phi p}$ are not identical in general.
Consequently, the $\phi A$ and $\bar\phi A$ scattering cross sections are different.
In the symmetric DM scenario, the average pointlike SI cross section of $\phi$ and $\bar\phi$ particles scattering off nuclei with mass number $A$ is given by
\begin{eqnarray}
\sigma_{\text{DM-}A}=\frac{\mu_{\phi A}^2}{2\pi}\{[Zf_{\phi p}+(A-Z)f_{\phi n}]^2+[Zf_{\bar{\phi} p}+(A-Z)f_{\bar{\phi} n}]^2\}.
\end{eqnarray}
Since 
\begin{eqnarray}
&& \frac{1}{2}\{ {[Z{f_{\phi p}} + (A - Z){f_{\phi n}}]^2} + {[Z{f_{\bar \phi p}} + (A - Z){f_{\bar \phi n}}]^2}\} 
\nonumber\\
 &=& {Z^2}\left[ {\frac{{{{(G_{\phi p}^{\mathrm{S}})}^2}}}{{4m_\phi ^2}} + {{(G_{\phi p}^{\mathrm{V}})}^2}} \right] + {(A - Z)^2}\frac{{{{(G_{\phi n}^{\mathrm{S}})}^2}}}{{4m_\phi ^2}} + 2Z(A - Z)\frac{{G_{\phi p}^{\mathrm{S}}G_{\phi n}^{\mathrm{S}}}}{{4m_\phi ^2}}
\end{eqnarray}
has no cross terms of the form $G^\mathrm{V}_{\phi N} G^\mathrm{S}_{\phi N}$, the interference between the vector current and scalar-type interactions actually cancels out for symmetric DM~\cite{Belanger:2013tla}.
For several isotopes $A_i$ with the same atomic number $Z$, the event rate per unit time in a direct detection experiment becomes
\begin{eqnarray}
R  &=& \frac{1}{{2\pi }}\sum\limits_i {{\eta _i}{I_{{A_i}}}\mu _{\phi {A_i}}^2\big\{ {{[Z{f_{\phi p}} + ({A_i} - Z){f_{\phi n}}]}^2} + {{[Z{f_{\bar \phi p}} + ({A_i} - Z){f_{\bar \phi n}}]}^2}\big\} } 
\label{R:CSDM}
\nonumber\\
&=& \frac{1}{2}{\sigma _{\phi p}}\sum\limits_i {{\eta _i}{I_{{A_i}}}\frac{{\mu _{\phi {A_i}}^2}}{{\mu _{\phi p}^2}}\bigg\{ {{{\left[ {Z + ({A_i} - Z)\frac{{{f_{\phi n}}}}{{{f_{\phi p}}}}} \right]}^2} + {{\left[ {Z\frac{{{f_{\bar \phi p}}}}{{{f_{\phi p}}}} + ({A_i} - Z)\frac{{{f_{\bar \phi n}}}}{{{f_{\phi p}}}}} \right]}^2}} \bigg\}}.
\end{eqnarray}

The experimental reports in terms of the normalized-to-nucleon cross section $\sigma_N^Z$ actually correspond to the assumption ${f_{\phi p}} = {f_{\bar \phi p}} = {f_{\phi n}} = {f_{\bar \phi n}}$, where the relation $\sigma_N^Z = \sigma_{\phi p}$ holds.
This leads to an expression similar to Eq.~\eqref{sigma_N_Z:IC},
\begin{equation}
\sigma _N^Z = \frac{R}{{\sum_i {{\eta _i}{I_{{A_i}}}A_i^2\mu _{\phi {A_i}}^2/\mu _{\phi p}^2} }}.
\end{equation}
In the realistic situation for our model, the above assumption is not satisfied, and the relation between $\sigma _N^Z$ and ${\sigma _{\phi p}}$ becomes
\begin{equation}\label{sigma_N_Z-sigma_phip}
\sigma _N^Z = {\sigma _{\phi p}}\,\frac{{\sum_i {{\eta _i}\mu _{\phi {A_i}}^2\big\{ {{[Z + ({A_i} - Z){f_{\phi n}}/{f_{\phi p}}]}^2} + {{[Z{f_{\bar \phi p}}/{f_{\phi p}} + ({A_i} - Z){f_{\bar \phi n}}/{f_{\phi p}}]}^2}\big\} } }}{{2\sum_i {{\eta _i}\mu _{\phi {A_i}}^2A_i^2} }}.
\end{equation}
Here, we have assumed that all $I_{A_i}$ are equal.

\begin{figure}[!t]
\centering
\subfigure[~~$\sigma_N^Z$\label{fig:sigma_Fi:sigma_N_Z}]
{\includegraphics[width=0.48\textwidth]{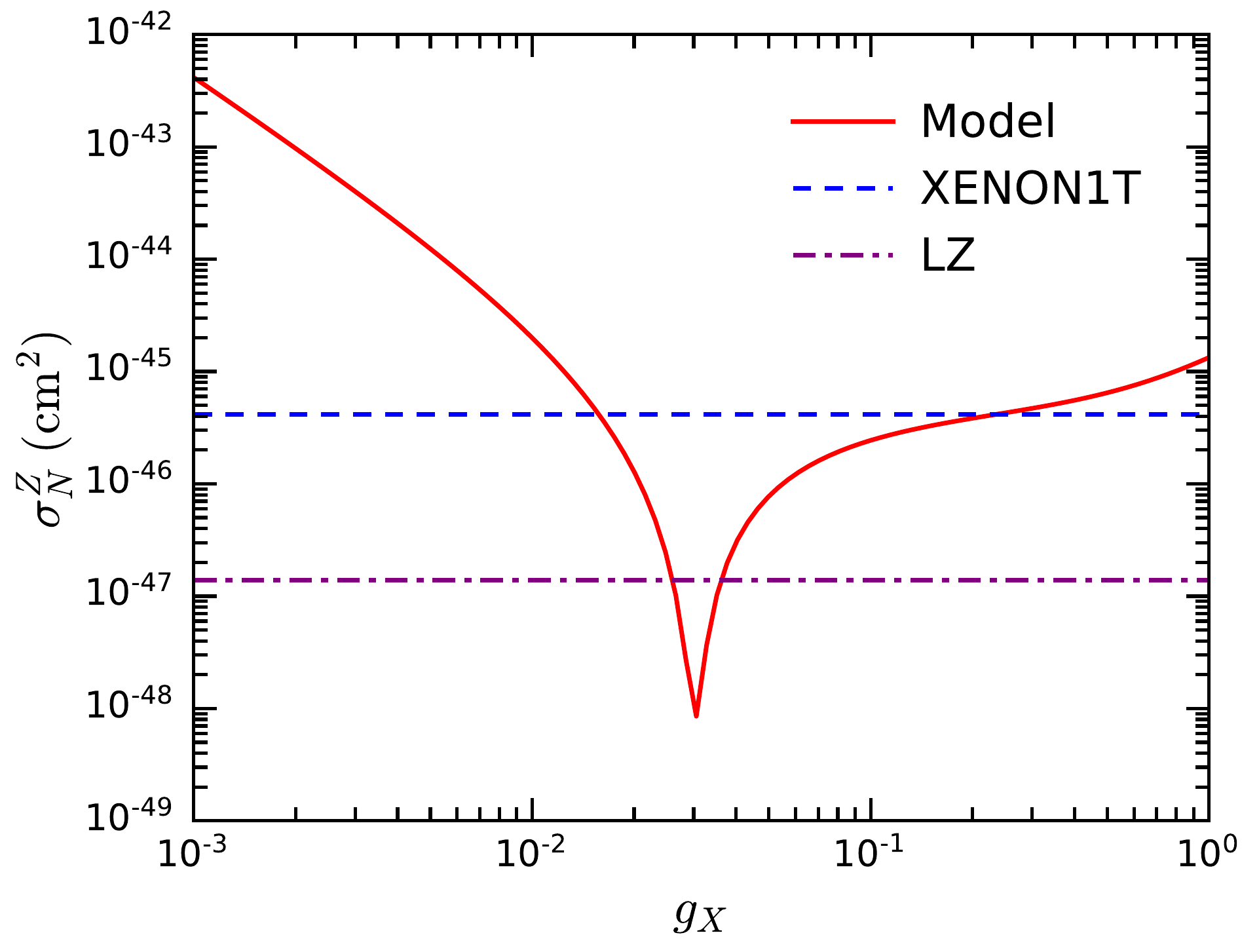}}
\hspace{1em}
\subfigure[~~$F_1$, $F_2$, and $F_1 + F_2$\label{fig:sigma_Fi:Fi}]
{\includegraphics[width=0.48\textwidth]{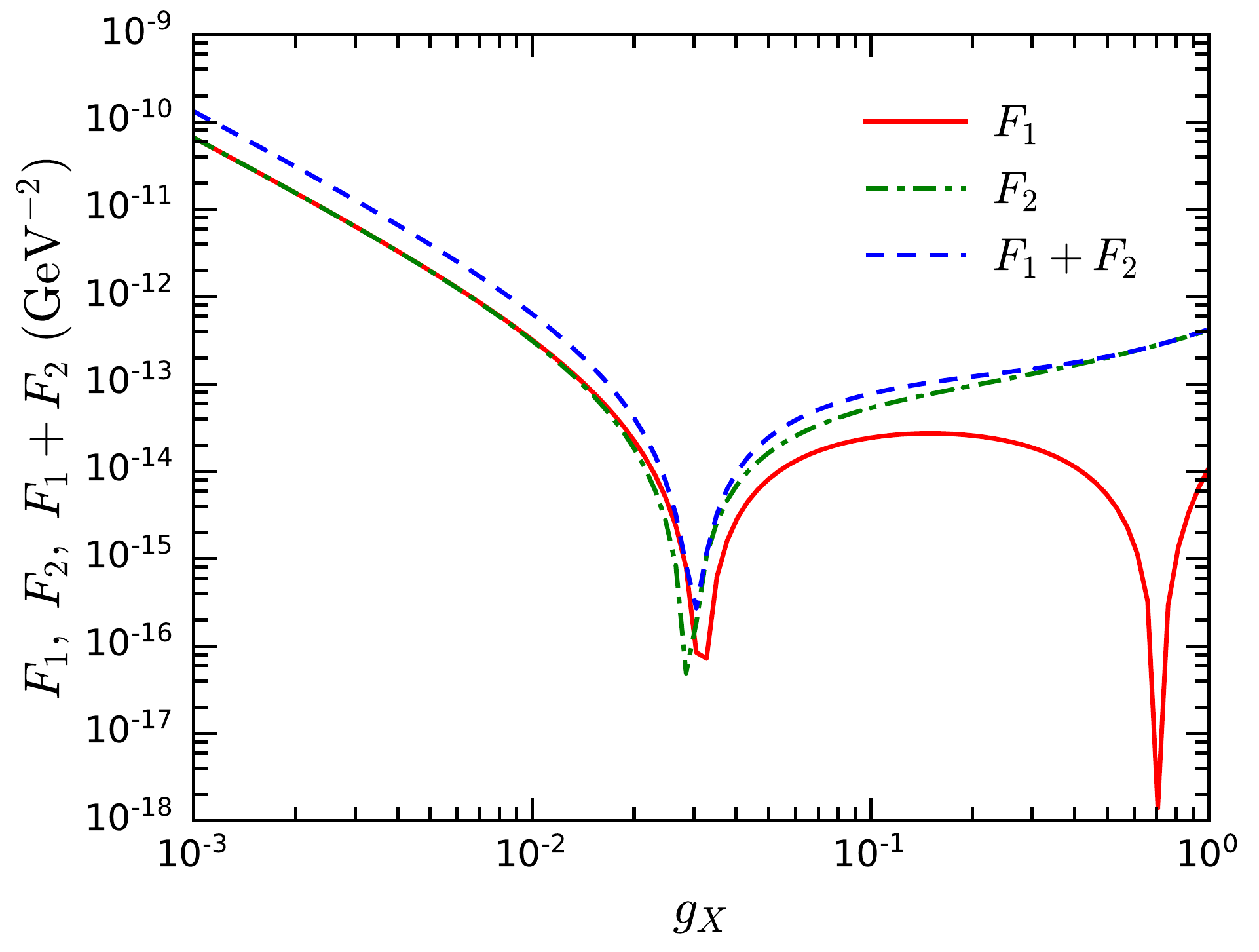}}
\caption{The normalized-to-nucleon cross section $\sigma_N^Z$ (a) and $F_1$, $F_2$, $F_1 + F_2$ (b) as functions of $g_X$ with the same fixed parameters in Fig.~\ref{fig:f_phiN} ($m_\phi=500~\si{GeV}$, $m_{Z'}=1~\si{TeV}$, $m_s=250~\si{GeV}$, $s_\varepsilon=0.1$, $s_\eta=0.01$, $\lambda_{H\phi }=0.1$, and $\lambda_{S\phi }=-0.1$).
The dashed blue line in the left panel denotes the 90\% C.L. upper bound on $\sigma_N^Z$ for $m_\phi=500~\si{GeV}$ from the XENON1T experiment~\cite{Aprile:2018dbl}.
The dot-dashed purple line indicates the 90\% C.L. sensitivity of the future LZ experiment~\cite{Mount:2017qzi}.}
\label{fig:sigma_Fi}
\end{figure}

In Fig.~\ref{fig:sigma_Fi:sigma_N_Z}, we display the DM-nucleon scattering cross section $\sigma_N^Z$ as a function of $g_X$ for the same fixed parameters adopted in Fig.~\ref{fig:f_phiN}.
For $g_X \lesssim 0.015$ and $g_X \gtrsim 0.22$, $\sigma_N^Z$ exceed the upper bound at $m_\phi = 500~\si{GeV}$ from the XENON1T experiment~\cite{Aprile:2018dbl}.
Nonetheless, there is a dip at $g_X \sim 0.03$, evading the XENON1T constraint and even the future LZ search.
We can understand this result through the following analysis.

The behavior of $\sigma_N^Z$ is essentially controlled by the two terms inside the curly bracket of the first line in Eq.~\eqref{R:CSDM}.
They can be approximately estimated by the following quantities:
\begin{equation}
{F_1} = {[Z{f_{\phi p}} + (\bar A - Z){f_{\phi n}}]^2},\quad
{F_2} = {[Z{f_{\bar \phi p}} + (\bar A - Z){f_{\bar \phi n}}]^2},
\end{equation}
where $\bar{A} = 131.293$ is the atomic weight for xenon.
Note that $F_1$ and $F_2$ are the contributions from the $\phi$ and $\bar\phi$ particles, respectively.
In Fig.~\ref{fig:sigma_Fi:Fi}, we show $F_1$, $F_2$, and their sum as functions of $g_X$.
We find that both the $F_1$ and $F_2$ curves have dips around $g_X \sim 0.03$,
because $f_{\phi p}$, $f_{\phi n}$, and $f_{\bar \phi p}$ are all close to zero around $g_X \sim 0.03$, as shown in Fig.~\ref{fig:f_phiN}.
The two dips lead to a dip at $g_X \sim 0.03$ in the $F_1 + F_2$ curve, explaining the dip in Fig.~\ref{fig:sigma_Fi:sigma_N_Z}.

Additionally, the $F_2$ curve has a second dip at $g_X
 \sim 0.7$.
The reason is that the ratio $f_{\phi n}/f_{\phi p}$ closes to $-Z/(\bar{A}-Z) \simeq -0.7$~\cite{Feng:2011vu} at $g_X \sim 0.7$, and the two terms inside the square bracket of the $F_2$ expression cancel each other out.
Nonetheless, this dip has no manifest effect in $F_1 + F_2$, since $F_2$ is much larger than $F_1$ at $g_X \sim 0.7$.
The $F_1 + F_2$ curve basically catches the behavior of $\sigma_N^Z$ in Fig.~\ref{fig:sigma_Fi:sigma_N_Z}.

\begin{figure}[!t]
	\centering
	\includegraphics[width=0.5\textwidth]{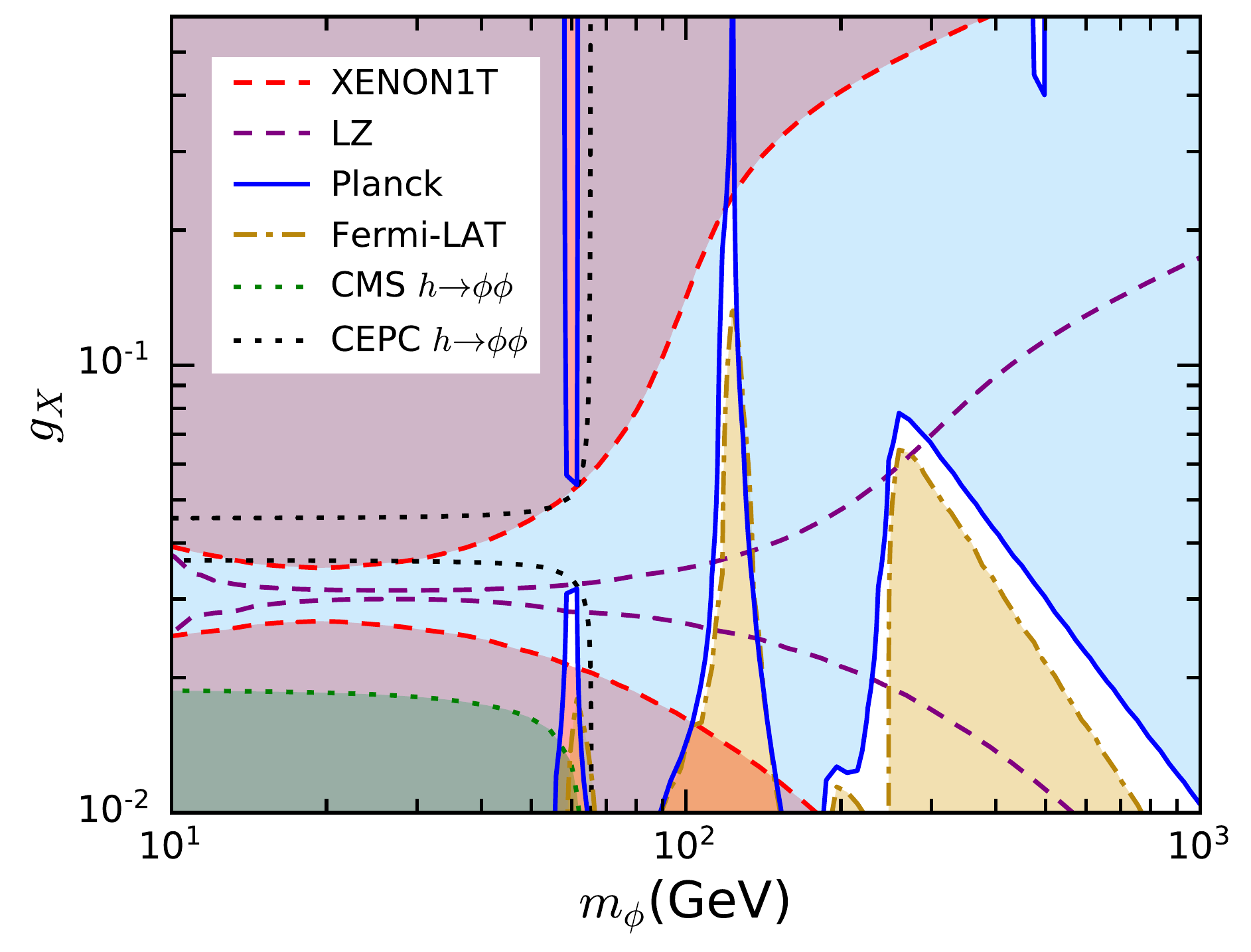}
	\caption{Experimental constraints from XENON1T, Planck, and Fermi-LAT in the $m_\phi$-$g_X$ plane for complex scalar DM with $m_{Z'}=1~\si{TeV}$, $m_s=250~\si{GeV}$, $s_\varepsilon=0.1$, $s_\eta=0.01$, $\lambda_{H\phi}=0.01$, and $\lambda_{S\phi}=-0.01$.
    The meanings of line types and colors are identical to those in Fig.~\ref{fig:mchi-gX:general}.
    In addition, the green shaded region is excluded by the CMS search for invisible Higgs decays~\cite{Khachatryan:2016whc}, while the dotted black lines denote the sensitivity of the future CEPC search for invisible Higgs decays~\cite{CEPCStudyGroup:2018ghi}.}
	\label{fig:mphi-gX:general}
\end{figure}

We utilize Eq.~\eqref{sigma_N_Z-sigma_phip} to derive the direct detection constraint.
In Fig.~\ref{fig:mphi-gX:general}, the red shaded areas are excluded at 90\% C.L. by the XENON1T experiment~\cite{Aprile:2018dbl} in the $m_\phi$-$g_X$ plane with fixed parameters $m_{Z'}=1000~\si{GeV}$, $m_s=250~\si{GeV}$, $s_\varepsilon=0.1$, $s_\eta=0.01$, $\lambda_{H\phi}=0.01$, and $\lambda_{S\phi}=-0.01$.
As discussed for Figs.~\ref{fig:f_phiN} and \ref{fig:sigma_Fi}, a region around $g_X \sim 0.03$ corresponds to a rather small $\sigma_N^Z$ and evades the XENON1T constraint.
Moreover, for $m_\phi \gtrsim 20~\si{GeV}$ the constraint becomes weaker and weaker as $m_\phi$ increases.
This is mainly because the $G^\mathrm{S}_{\phi N}/(2m_\phi)$ terms in $f_{\phi N}$ and $f_{\bar\phi N}$ are suppressed by $m_\phi$.
The future LZ experiment will probe much larger regions than XENON1T does.

\subsection{Relic abundance and indirect detection}

Now we discuss the constraints from relic abundance observation and indirect detection.
Analogous to Dirac fermionic DM, the possible $\phi\bar\phi$ annihilation channels include $f\bar{f}$, $W^+W^-$, $h_i h_j$, $Z_iZ_j$, and $h_iZ_j$, with $h_i\in\{h,s\}$ and $Z_i \in \{Z,Z'\}$.
Nonetheless, these annihilation processes are not only induced by the kinetic mixing portal, but also by the Higgs portal.
In Fig.~\ref{fig:mphi-gX:general}, the solid blue lines correspond to the correct relic abundance, while the blue shaded areas predict DM overproduction.
The orange shaded areas are excluded at 95\% C.L. by the Fermi-LAT experiment~\cite{Ackermann:2015zua}.

There are several available regions for the relic abundance observation.
Firstly, two available strips around $m_\phi \sim m_h/2 = 62.5~\si{GeV}$ are related to resonant annihilation at the $h$ pole.
These strips cannot meet each other because the $h\phi\phi$ coupling $({\lambda _{S\phi }}{s_\eta }{v_S} + {\lambda _{H\phi }}{c_\eta }v)$ approaches zero at $g_X \sim 0.04$.
Nonetheless, the upper strip is excluded by XENON1T, while a section of the lower strip is free from current experimental constraints but may be tested by LZ.

In addition, both the $ZZ$ annihilation channel opening for $m_\chi \gtrsim m_Z$ and the resonance of the $s$ boson at $m_\phi \sim m_s/2 =125~\si{GeV}$ contribute to a narrow available region with $90~\si{GeV} \lesssim m_\phi \lesssim 150~\si{GeV}$.
Only a small fraction of this region evades the constraints from XENON1T and Fermi-LAT.
Moreover, a broad available region with $170~\si{GeV} \lesssim m_\phi \lesssim 1~\si{TeV}$ is contributed by the $s Z$ and $ss$ annihilation channels opening for $m_\phi \gtrsim 170~\si{GeV}$ and $m_\phi \gtrsim 250~\si{GeV}$, respectively.
This region circumvents the XENON1T constraint but faces the Fermi-LAT constraint.
Note that the LZ experiment will further explore  these two regions.

The annihilation processes contributing to the above available regions are primarily induced by the Higgs portal.
Nonetheless, there is another available strip with $g_X \gtrsim 0.4$ at $m_\phi \sim m_{Z'}/2 = 500~\si{GeV}$ corresponding to the resonant annihilation at the $Z'$ pole, which is induced by the $\mathrm{U}(1)_\mathrm{X}$ gauge interaction and the kinetic mixing portal.
For $g_X < 0.6$, this strip is free from the direct and indirect detection constraints.

\begin{figure}[!t]
	\centering
	\subfigure[~~$30~\si{GeV}\leq m_{Z'} \leq 60~\si{GeV}$\label{fig:mphi-mZp:light_Zp}]
	{\includegraphics[width=.46\textwidth]{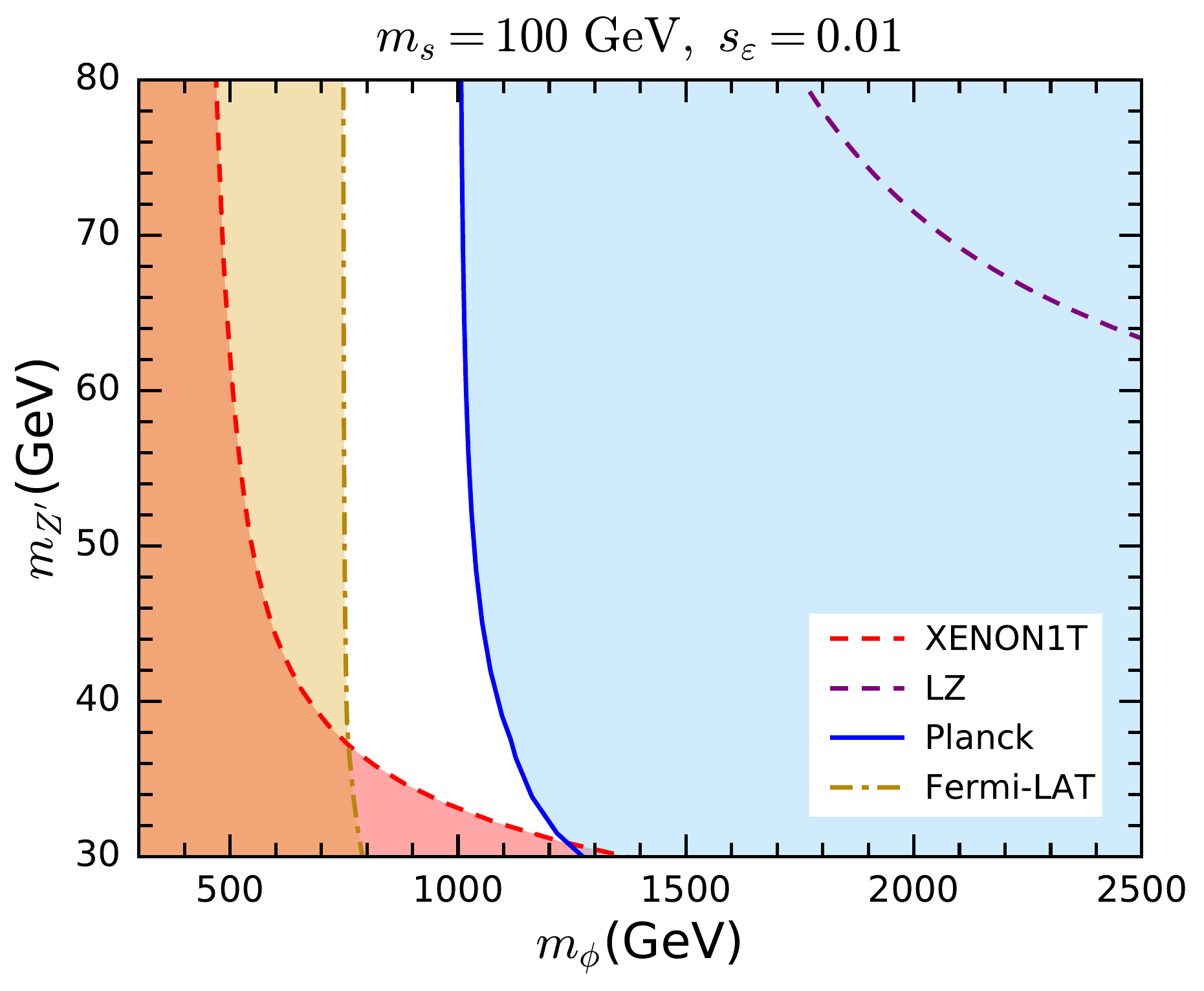}}
	\subfigure[~~$400~\si{GeV} \leq m_{Z'} \leq 1.5~\si{TeV}$\label{fig:mphi-mZp:heavy_Zp}]
	{\includegraphics[width=.48\textwidth]{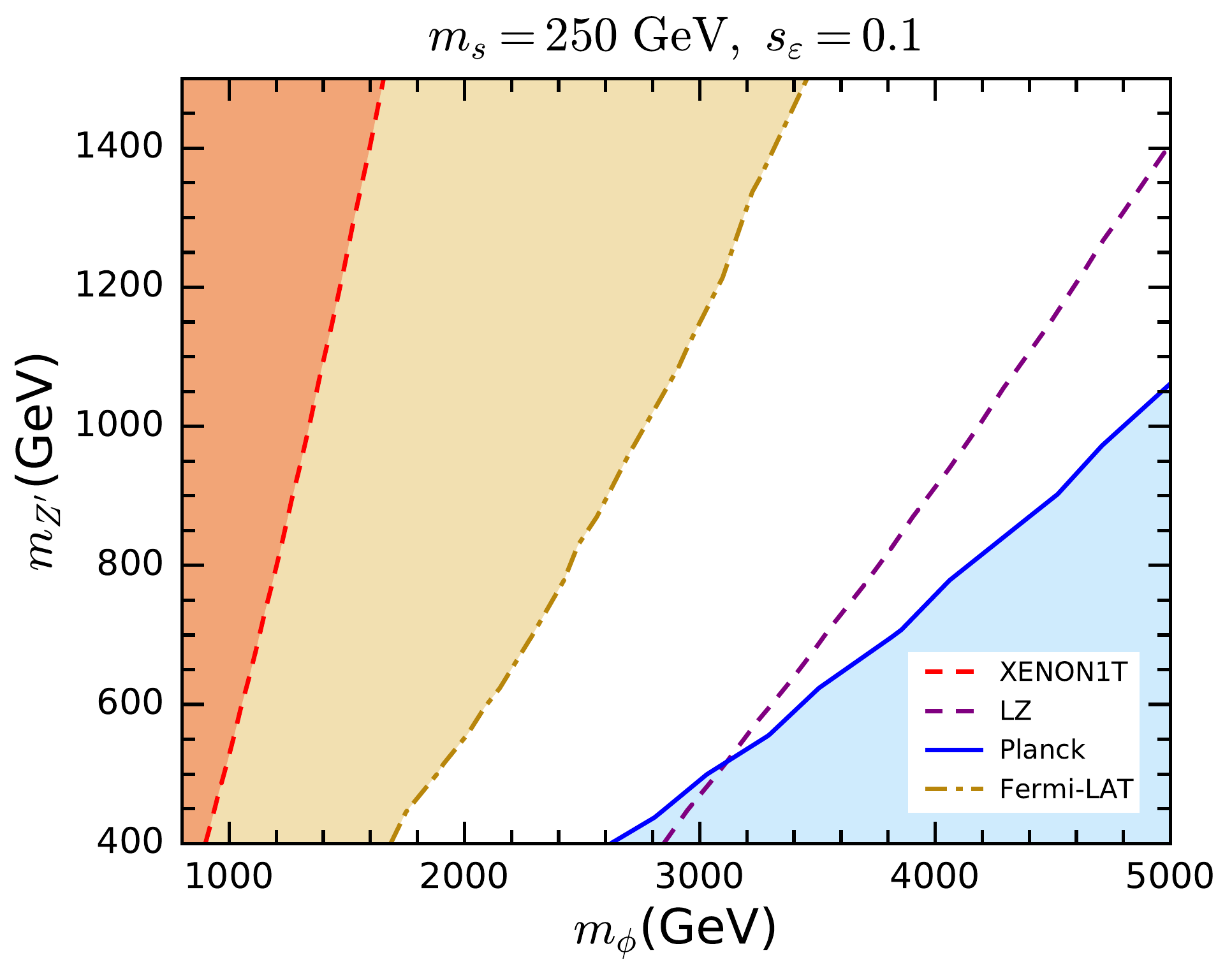}}
	\caption{Experimental constraints in the $m_\phi$-$m_{Z'}$ plane for complex scalar DM in the $30~\si{GeV}\leq m_{Z'} \leq 60~\si{GeV}$ range with $m_s=100~\si{GeV}$ and $s_{\varepsilon}=0.01$ (a) and in the $400~\si{GeV} \leq m_{Z'} \leq 1.5~\si{TeV}$ range with $m_s=250~\si{GeV}$ and $s_{\varepsilon}=0.1$ (b).
    Other parameters in both panels are fixed as $g_X=0.01$, $s_\eta=0.01$, and $\lambda_{H\phi}=\lambda_{S\phi}=0.1$.}
	\label{fig:mphi-mZp}
\end{figure}

Below we study the phenomenology in the planes of other parameter pairs.
The experimental constraints in the $m_\phi$-$m_{Z'}$ plane are demonstrated in the two panels of Fig.~\ref{fig:mphi-mZp} for $g_X=0.01$, $s_\eta=0.01$, and $\lambda_{H\phi}=\lambda_{S\phi}=0.1$.
In Fig.~\ref{fig:mphi-mZp:light_Zp} with $m_s=100~\si{GeV}$ and $s_{\varepsilon}=0.01$, $Z'$ is light ($30~\si{GeV}\leq m_{Z'} \leq 60~\si{GeV}$), and the vector current interactions are dominant in DM-nucleus scattering.
Therefore, the XENON1T bound is more stringent for lighter $Z'$, excluding up to $m_\phi \sim 1.35~\si{TeV}$ at $m_{Z'} = 30~\si{GeV}$.
The correct relic abundance is corresponding to a curve with $m_\phi \sim 1 \text{--} 1.3~\si{TeV}$, while the Fermi-LAT experiment excludes a region with $m_\phi \lesssim 800~\si{GeV}$.
The region survived from the above constraints will be covered by the LZ experiment.

On the other hand, $Z'$ is heavy ($400~\si{GeV} \leq m_{Z'} \leq 1.5~\si{TeV}$) in Fig.~\ref{fig:mphi-mZp:heavy_Zp} with $m_s=250~\si{GeV}$ and $s_{\varepsilon}=0.1$, and thus, the scalar-type interactions are important in direct detection.
Because $g_X$ is fixed, $v_S$ increases with $m_{Z'}$ following Eq.~\eqref{v_S}.
As a result, the XENON1T constraint is stricter for heavier $Z'$, excluding up to $m_\phi \sim 1.65~\si{TeV}$ at $m_{Z'} = 1.5~\si{TeV}$.
In this case, the Fermi-LAT constraint is even more stringent, ruling out a region with $m_\phi \lesssim 3.45~\si{TeV}$.
The observed relic abundance corresponds to a curve with $m_\phi \gtrsim 2.5~\si{TeV}$, which is not excluded by XENON1T but will be tested by LZ for $m_\phi \lesssim 3.1~\si{TeV}$.

\begin{figure}[!t]
	\centering
	\subfigure[~~$m_\phi$-$\lambda_{S\phi}$ plane\label{fig:lam_Sphi:mphi}]
	{\includegraphics[width=.48\textwidth]{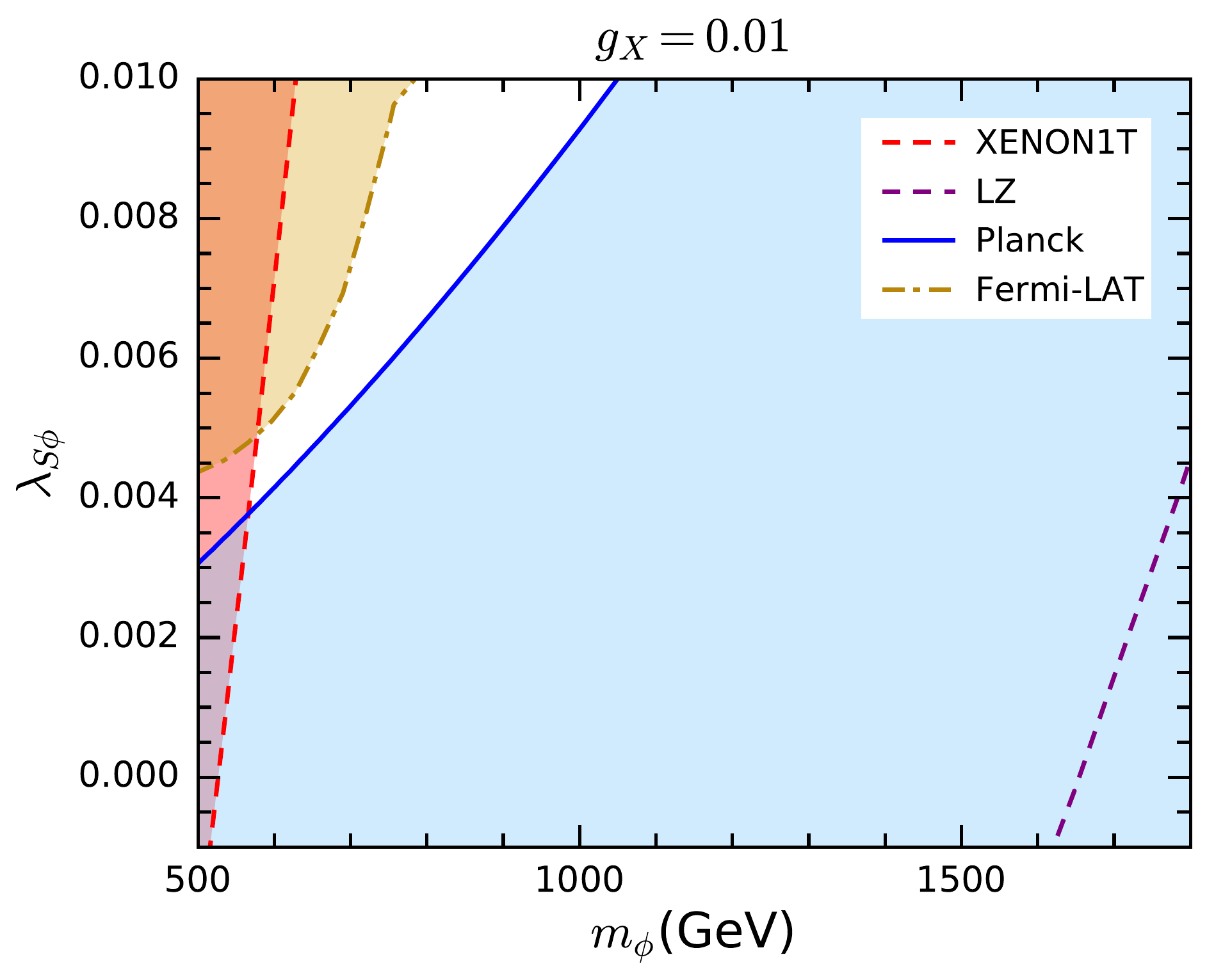}}
	\subfigure[~~$g_X$-$\lambda_{S\phi}$ plane\label{fig:lam_Sphi:gX}]
	{\includegraphics[width=.472\textwidth]{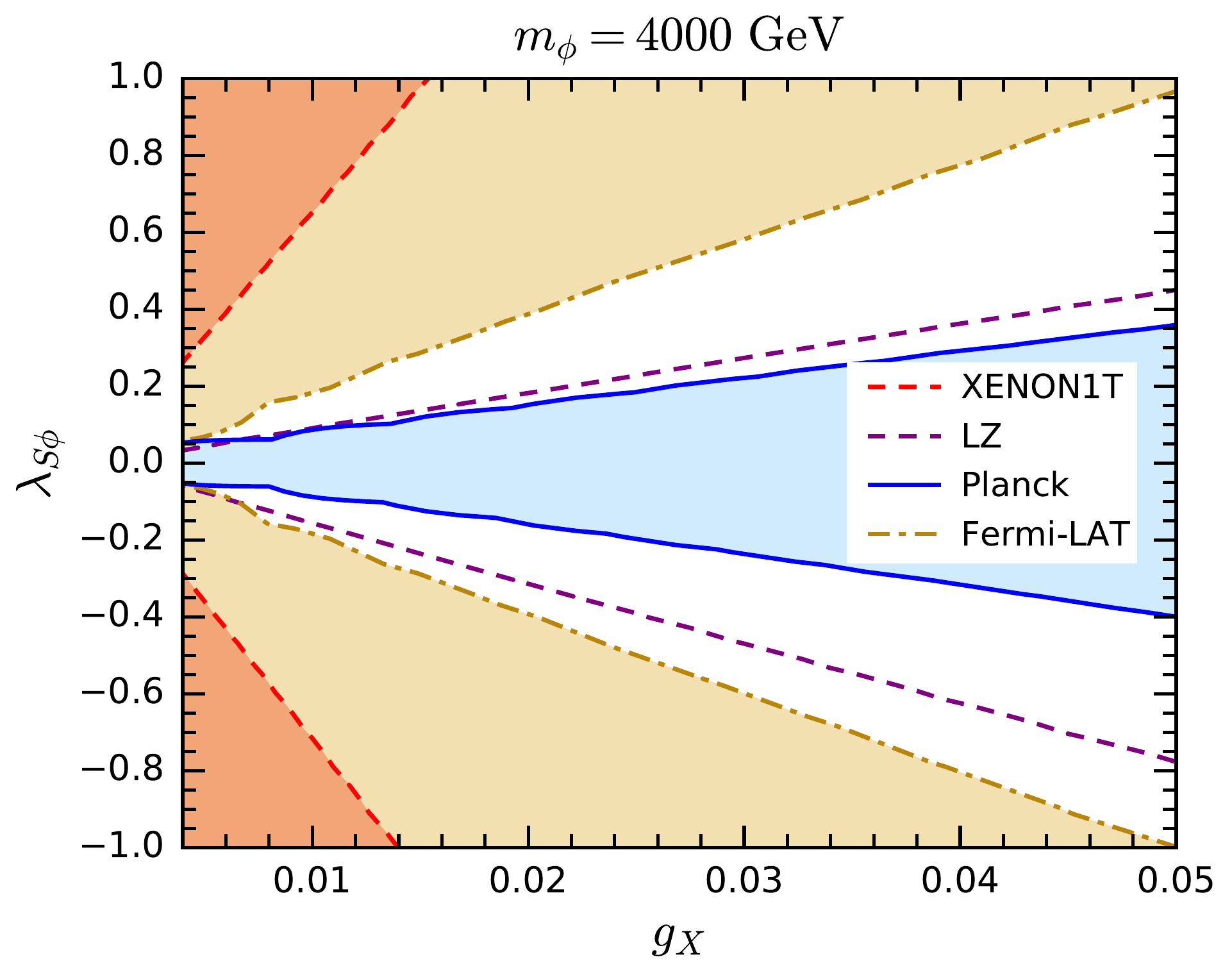}}
	\caption{Experimental constraints in the $m_\phi$-$\lambda_{S\phi}$ plane for complex scalar DM with $g_X = 0.01$  (a) and in the $g_X$-$\lambda_{S\phi}$ plane with $m_\phi = 4~\si{TeV}$ (b).
    The common parameters in both panels are $m_{Z'}=1~\si{TeV}$, $m_s=250~\si{GeV}$, $s_\varepsilon=0.1$, $s_\eta=0.01$, and $\lambda_{H\phi}=0.1$.}
	\label{fig:lam_Sphi}
\end{figure}

The experimental constraints are also displayed in $m_\phi$-$\lambda_{S\phi}$ plane with $g_X = 0.01$ in Fig.~\ref{fig:lam_Sphi:mphi}, as well as in the $g_X$-$\lambda_{S\phi}$ plane with $m_\phi = 4~\si{TeV}$ in Fig.~\ref{fig:lam_Sphi:gX}.
The other parameters in both plots are fixed as $m_{Z'}=1~\si{TeV}$, $m_s=250~\si{GeV}$, $s_\varepsilon=0.1$, $s_\eta=0.01$, and $\lambda_{H\phi}=0.1$.
In Fig.~\ref{fig:lam_Sphi:mphi}, the relic abundance observation is corresponding to a curve with $0.0032 \lesssim \lambda_{S\phi} \lesssim 0.0067$ in the range of $500~\si{GeV} \leq m_\phi \leq 800~\si{GeV}$.
This curve totally evades the Fermi-LAT constraint but is excluded for $m_\phi \lesssim 570~\si{GeV}$ by XENON1T.
The LZ experiment will test the whole curve.
In Fig.~\ref{fig:lam_Sphi:gX}, the correct relic abundance corresponds to two curves with $0.05 \lesssim \lambda_{S\phi} \lesssim 0.35$ and $-0.4 \lesssim \lambda_{S\phi} \lesssim -0.05$ in the range of $0.004 \leq g_X \leq 0.05$.
Both the direct and indirect detection experiments cannot exclude these two curves.

\subsection{Invisible Higgs decays}

If $m_\phi < m_h/2$, the $h\to\phi\phi$ decay is allowed.
Since detectors at colliders are generally unable to measure DM particles, the $\phi\phi$ final state is invisible, typically giving rise to signatures with missing transverse momentum.
In other words, $h\to\phi\phi$ is an invisible Higgs decay process.

From the interaction Lagrangian \eqref{eq:L_phi_h_s}, we derive the partial decay width of $h\to\phi\phi$ as 
\begin{equation}
\Gamma _{{\mathrm{inv}}}^h = \frac{{{{({\lambda _{H\phi }}{c_\eta }v + {\lambda _{S\phi }}{s_\eta }{v_S})}^2}}}{{16\pi {m_h}}}\sqrt {1 - \frac{{4m_\phi ^2}}{{m_h^2}}} \,.
\end{equation}
Since Eq.~\eqref{eq:HS_to_hs} leads to $H = c_\eta h - s_\eta s$, the $h$ couplings to $W$ and SM fermions just deviate from the corresponding couplings in the SM by a factor of $c_\eta$.
Accordingly, the partial widths of $h$ decays into $f\bar{f}$, $W^+ W^-$, and $gg$ are scaled with a factor of $c_\eta^2$.
The $h\to ZZ$ decay width would also depend on other parameters, but its contribution to the total decay width ${\Gamma ^h}$ is small.
Therefore, we have a good approximate relation ${\Gamma ^h} \simeq c_\eta ^2\Gamma _{{\mathrm{SM}}}^h$, where ${\Gamma _{{\mathrm{SM}}}^h} = 4.07~\si{MeV}$~\cite{Tanabashi:2018oca} is the total decay width of the Higgs boson in the SM.
Thus, the branching ratio of invisible Higgs decays in our model can be expressed as
\begin{equation}
{\mathcal{B}_{{\mathrm{inv}}}} \simeq \frac{{\Gamma _{{\mathrm{inv}}}^h}}{{c_\eta ^2\Gamma _{{\mathrm{SM}}}^h + \Gamma _{{\mathrm{inv}}}^h}}.
\end{equation}

The CMS search for invisible Higgs decays combining the $7$, $8$, and $13~\si{TeV}$ LHC data gives a bound of $\mathcal{B}_\mathrm{inv} < 24\%$ at 95\% C.L.~\cite{Khachatryan:2016whc}.
Such a bound can be used to constrain the parameter space for $m_\phi < m_h/2$.
We overlay this constraint in Fig.~\ref{fig:mphi-gX:general}, finding that it is weaker than the XENON1T constraint.

Future Higgs factories, like CEPC and FCC-ee, would be extremely sensitive to invisible Higgs decays.
The 95\% C.L. projected CEPC sensitivity for a data set of $5.6~\si{fb^{-1}}$ is $\mathcal{B}_\mathrm{inv} < 0.3\%$~\cite{CEPCStudyGroup:2018ghi}.
FCC-ee is expected to reach comparable sensitivity~\cite{Abada:2019zxq}.
Expressing the CEPC sensitivity in Fig.~\ref{fig:mphi-gX:general}, we find that CEPC would efficiently explore the parameter regions with $m_\phi < m_h/2$, except for a narrow zone around $g_X \sim 0.04$, where the $h\phi\phi$ coupling is close to zero.
Note that the CEPC search could probe the survived strip with $m_\phi \sim 60~\si{GeV}$ and $g_X \sim 0.02 \text{--} 0.03$.

\section{Conclusions and discussions}
\label{sec:concl}

In this work, we have explored the phenomenology of Dirac fermionic and complex scalar DM with hidden $\mathrm{U}(1)_\mathrm{X}$ gauge interaction and kinetic mixing between the $\mathrm{U}(1)_\mathrm{X}$ and $\mathrm{U}(1)_\mathrm{Y}$ gauge fields.
Besides the DM particle, the extra particles beyond the SM involve a massive neutral vector boson $Z'$ and a Higgs boson $s$ originated from the Brout-Englert-Higgs mechanism that gives mass to the $\mathrm{U}(1)_\mathrm{X}$ gauge field.
The measurement of the electroweak oblique parameters $S$ and $T$ puts a stringent constraint on the kinetic mixing parameter $s_\varepsilon$ if $Z'$ is not too heavy. 

For the Dirac fermionic DM particle $\chi$, the kinetic mixing term provides a portal for interactions with SM fermions, inducing potential signals in DM direct and indirect detection experiments.
In such a case, the DM-nucleon interactions are isospin violating.
More specifically, $\chi$ scatters off protons, but not off neutrons at the zero momentum transfer limit.
This leads to weaker direct detection constraints than those under the conventional assumption of isospin conservation.

Assuming DM is thermal produced in the early Universe, we
have investigated the parameter regions that are consistent with the relic abundance observation.
As the kinetic mixing parameter $s_\varepsilon$ has been bounded to be small, the available regions arise from the resonant annihilation at the $Z'$ pole or the $sZ'$ annihilation channel with dark sector interactions.
These regions have not been totally explored in the XENON1T direct detection and Fermi-LAT indirect detection experiments.
The future LZ experiment will investigate the parameter space much further.

For the complex scalar DM particle $\phi$, the communications with SM particles are not only through the kinetic mixing portal, but also through the Higgs portal arising from the scalar couplings.
The DM-nucleon scattering is still isospin violating.
Moreover, the $\bar\phi p$ scattering cross section is typically different from the $\phi p$ scattering cross section.
After a dedicated analysis, we have found that the XENON1T constraint can be significantly relaxed for particular parameters that leads to a cancellation effect between the $h$ and $s$ propagators.

For the relic abundance observation, our calculation has shown that there are several available regions, corresponding to the resonant annihilation at the $h$, $s$, and $Z'$ poles, as well as the $ZZ$, $sZ$, and $ss$ annihilation channels.
Additionally, we have carried out further investigations in the parameter space.
We have found that there are still a lot of parameter regions that predict an observed relic abundance but have not been excluded by the direct and indirect detection experiments.
The LZ experiment will provide further tests for these regions.
If $m_\phi < m_h/2$, the $h\to \phi\phi$ decay is allowed, and searches for invisible Higgs decays at a future Higgs factory will be rather sensitive.

An important difference between the Dirac fermionic and complex scalar DM models is that $\chi$ and $\phi$ have different spins.
Spin determination would be crucial for distinguishing various DM models once the DM particle is discovered.
Utilizing the angular distribution of nuclear recoils, a study in Ref.~\cite{Catena:2017wzu} showed that $\sim 100$ signal events in next generation directional direct detection experiments could be sufficient to distinguish spin-0 DM (like $\phi$) from spin-1/2 (like $\chi$) or spin-1 DM.

\begin{acknowledgments}

The authors acknowledge Yi-Lei Tang for helpful discussions.
This work is supported in part by the National Natural Science Foundation of China under Grants No.~11805288, No.~11875327, and No.~11905300, the China Postdoctoral Science Foundation
under Grant No.~2018M643282, the Natural Science Foundation of Guangdong Province
under Grant No.~2016A030313313,
the Fundamental Research Funds for the Central Universities,
and the Sun Yat-Sen University Science Foundation.

\end{acknowledgments}

\appendix

\section{Parameter relations}
\label{param_relation}

In Sec.~\ref{sec:U1X_gauge}, we choose a set of independent parameters $\{g_X, m_{Z'}, m_s,  s_{\varepsilon}, s_{\eta}\}$, from which other parameters can be derived.

Utilizing Eqs.~\eqref{massofgauge:Z}, \eqref{massofgauge:Zprime}, and \eqref{r}, we can derive a quadratic equation for $t_\xi$ from Eq.~\eqref{t2xi},
\begin{equation}\label{eq:t_xi}
{{\hat s}_{\mathrm{W}}}{t_\varepsilon }rt_\xi ^2 + (r - 1){t_\xi } + {{\hat s}_{\mathrm{W}}}{t_\varepsilon } = 0.
\end{equation}
The physical solution is
\begin{eqnarray}
t_{\xi} = \frac{2\hat{s}_\mathrm{W}t_\varepsilon}{1-r}
\left[1+\sqrt{1-r\left(\frac{2\hat{s}_\mathrm{W}t_\varepsilon}{1-r}\right)^2}\right]^{-1}.
\label{t_xi}
\end{eqnarray}
If $t_\varepsilon \neq 0$, there is no solution for $r=1$ (i.e., $m_{Z'} = m_Z$).
For $r \neq 1$, the solution exists only if the condition $[2\hat{s}_\mathrm{W}t_\varepsilon/(1-r)]^2\leq 1/r$ is satisfied.
For a small $t_\varepsilon$, we have $t_\xi \simeq \hat{s}_\mathrm{W} t_\varepsilon /(1-r)$~\cite{Arcadi:2018tly}.

With the solution \eqref{t_xi}, we can numerically solve Eq.~\eqref{weak_mixing_angle} and obtain $\hat{s}_\mathrm{W}$ as a function of $s_\varepsilon$ and $m_{Z'}$.
Then $t_\xi$ is also a function of $s_\varepsilon$ and $m_{Z'}$.

From Eq.~\eqref{massofgauge:Zprime}, we can derive the VEV of the hidden Higgs field as
\begin{equation}\label{v_S}
v_{S}=\frac{m_{Z^{\prime}} c_{\varepsilon}}{g_{X}} \sqrt{1+\hat{s}_\mathrm{W} t_{\varepsilon} t_{\xi}}\,.
\end{equation}
Because of Eqs.~\eqref{m_h_2}, \eqref{m_s_2}, and \eqref{t_2eta}, the scalar quartic couplings are given by
\begin{eqnarray}
\lambda_H &=&\frac{(m_s^2+m_h^2)-c_{2\eta}(m_s^2-m_h^2)}{2v^2},\\
\lambda_S &=&\frac{(m_s^2+m_h^2)+c_{2\eta}(m_s^2-m_h^2)}{2v^2},\\
\lambda_{HS} &=&\frac{t_{2\eta}(\lambda_Hv^2-\lambda_Sv_S^2)}{2vv_S}.
\end{eqnarray}

\bibliographystyle{utphys}
\bibliography{reference}

\end{document}